\documentclass[final,12pt]{msml2021}
\usepackage{amsfonts}
\usepackage{algorithm, algorithmic}

\newcommand{\ATLNote}[1]{}

\newcommand{\edit}[1]{#1}

\renewcommand{\cite}{\citep}

\title[Decentralization by Imitation of Centralized Controller]{Decentralized Multi-Agents by Imitation of a Centralized Controller}
\usepackage{times}

\msmlauthor{%
 \Name{Alex Tong Lin} \Email{atlin@math.ucla.edu}\\
 \addr University of California, Los Angeles
 \AND
 \Name{Mark J. Debord} \Email{}\\
 \addr NAVAIR%
 \AND
 \Name{Katia Estabridis} \Email{} \\
 \addr NAVAIR%
 \AND 
 \Name{Gary Hewer} \Email{} \\
 \addr NAVAIR%
 \AND 
 \Name{Guido Mont\'{u}far} \Email{montufar@math.ucla.edu} \\
 \addr University of California, Los Angeles, and Max Planck Institute, Leipzig
 \AND 
 \Name{Stanley Osher} \Email{sjo@math.ucla.edu} \\
 \addr University of California, Los Angeles
}

\makeatletter
 \let\Ginclude@graphics\@org@Ginclude@graphics
\makeatother

\begin{document}

\maketitle

\begin{abstract}%
  We consider a multi-agent reinforcement learning problem where each agent seeks to maximize a shared reward while interacting with other agents, and they may or may not be able to communicate. Typically the agents do not have access to other agent policies and thus each agent is situated in a non-stationary and partially-observable environment. In order to obtain multi-agents that act in a decentralized manner, we introduce a novel algorithm under the popular framework of centralized training, but decentralized execution. This training framework first obtains solutions to a multi-agent problem with a single centralized joint-space learner, which is then used to guide imitation learning for independent decentralized multi-agents. This framework has the flexibility to use any reinforcement learning algorithm to obtain the expert as well as any imitation learning algorithm to obtain the decentralized agents. This is in contrast to other multi-agent learning algorithms that, for example, can require more specific structures. We present some theoretical bounds for our method, and we show that one can obtain decentralized solutions to a multi-agent problem through imitation learning.
\end{abstract}


\section{Introduction}

Reinforcement Learning (RL) is the problem of finding an action policy that maximizes reward for an agent embedded in an environment \cite{SuttonB18}. It has recently has seen an explosion in popularity due to its many achievements in various fields such as robotics \cite{levine2016end}, industrial applications \cite{datacentercooling}, game-playing \cite{mnih2015human, silver2017mastering, silver2016mastering}, and the list continues. However, most of these achievements have taken place in the single-agent realm, where one does not have to consider the dynamic environment provided by interacting agents that learn and affect one another. 

This is the problem of Multi-agent Reinforcement Learning (MARL) where we seek to find the best action policy for each agent in order to maximize their reward. The settings may be cooperative, and thus they might have a shared reward, or the setting may be competitive, where one agent's gain is another's loss. Some examples of a multi-agent reinforcement learning problem are: decentralized coordination of vehicles to their respective destinations while avoiding collision, or the game of pursuit and evasion where the pursuer seeks to minimize the distance between itself and the evader while the evader seeks the opposite. Other examples of multi-agent tasks can be found in \cite{Panait2005} and~\cite{maddpg}. 

The key difference between MARL and single-agent RL (SARL) is that of interacting agents, which is why the achievements of SARL cannot be absentmindedly transferred to find success in MARL. Specifically, the state transition probabilities in a MARL setting are inherently non-stationary from the perspective of any individual agent. This is due to the fact that the other agents in the environment are also updating their policies, and so the Markov assumptions typically needed for SARL convergence are violated. This aspect of MARL gives rise to instability during training, where each agent is essentially trying to learn a moving target.

In this work, we present a novel method for MARL in the cooperative setting (with shared reward). Following the popular framwork of centralized training/learning but decentralized execution, our method first trains a centralized expert with full observability, and then uses this expert as a supervisor for independently learning agents. There are a myriad of imitation/supervised learning algorithms, and in this work we focus on adapting DAgger (Dataset Aggregation) \cite{DAgger} to the multi-agent setting. After the imitation learning stage, the agents are able to successfully act in a decentralized manner. We call this algorithm Centralized Expert Supervises Multi-Agents (CESMA). CESMA adopts the framework of centralized training, but decentralized execution \cite{KRAEMER201682}, the end goal of which is to obtain multi-agents that can act in a decentralized manner.

\section{Related works}

The most straight-forward way of adapting single-agent RL algorithms to the multi-agent setting is by having agents be independent learners. This was applied in \cite{Tan:1997:MRL:284860.284934}, but this training method gives instability issues, as the environment is non-stationary from the perspective of each agent \cite{Matignon:2012:RIR:2349641.2349642, Busoniu2010MultiagentRL, claus1998dynamics}. This non-stationarity was examined in \cite{OmidshafieiPAHV17}, and stabilizing experience replay was studied in \cite{foerster2017stabilising}. 

Another common approach to stabilizing the environment is to allow the multi-agents to communicate. In \cite{SukhbaatarSF16}, they examine this using continuous communications so one may backpropagate to learn to communicate. And in \cite{foerster2016learning}, they give an in-depth study of communicating multi-agents, and also provide training methods for discrete communication. In \cite{jpaulos_schen}, they decentralize a policy by examining what to communicate and by utilizing supervised learning, although they mathematically solve for a centralized policy and their assumptions require homogeneous communicating agents. 

Others approach the non-stationarity issue by having the agents take turns updating their weights while freezing others for a time, although non-stationarity is still present \cite{egorov2016multi}. Other attempts adapt $Q$-learning to the multi-agent setting: Distributed $Q$-Learning \cite{Lauer00analgorithm} updates $Q$-values only when they increase, and updates the policy only for actions that are not greedy with respect to the $Q$-values, and Hysteretic $Q$-Learning \cite{hysteretic} provides a modification. Other approaches examine the use of parameter sharing \cite{gupta2017cooperative} between agents, but this requires a degree of homogeneity of the agents. And in \cite{tesauro2004extending}, their approach to non-stationarity was to input other agents' parameters into the $Q$ function. Other approaches to stabilize the training of multi-agents are in \cite{sukhbaatar2016learning}, where the agents share information before selecting their actions. There is also recently CoDAIL \cite{codail}, where they use reasonable assumptions on the the correlation between policies, and in \cite{NIPS2019_8402} they examine multi-agents that are of nodes of a network and where each agent can only communicate with its neighbor. A population-based training regime based on game-theoretic ideas called Policy-Spaced Response Oracles (PSRO) \cite{lanctot2017unified} is given in \cite{Muller2020A}.

From a more centralized view point, \cite{oliehoek2008optimal, qmix, sunehag2017value} derived a centralized $Q$-value function for MARL, and in \cite{DBLP:journals/corr/UsunierSLC16}, they train a centralized controller and then sequentially select actions for each agent. The issue of an exploding action space was examined in \cite{tavakoli2018action}.

A few works that follow the framework of centralized training, but decentralized execution are: RLar (Reinforcement Learning as Rehearsal) \cite{KRAEMER201682}, COMA (Counterfactual Multi-Agent), and also \cite{ijcai2018-774, DBLP:journals/corr/abs-1810-11702} -- where the idea of knowledge-reuse is examined. In \cite{NIPS2017_6887}, they examine decentralization of policies from an information-theoretic perspective. There is also MADDPG \cite{maddpg}, where they train in a centralized-critics decentralized-actors framework; after training completes, the agents are separated from the critics and can execute in a fully distributed manner. 

In the flavor of inverse reinforcement learning and imitation learning, there is \cite{le2017coordinated}, where they incorporate structure learning with conventional imitation learning. And in MA-AIRL \cite{maairl}, where they apply ideas from inverse reinforcement learning in order to discover the reward policies for multi-agents, and similarly there is MA-GAIL \cite{magail}, where they use a generative and adversarial framework in order to discover the proper reward functions.

For surveys of MARL, see articles in \cite{bu2008comprehensive, panait2005cooperative}. \edit{For a survey of imitation learning, the interested reader can see \cite{osa2018algorithmic}.}

\section{Background}

In this section we briefly review the requisite material needed to define MARL problems. Additionally we summarize some of the standard approaches in general reinforcement learning and discuss their use in MARL.

\textbf{Dec-POMDP:} A formal framework for multi-agent systems is called a decentralized partially-observable Markov decision process (Dec-POMDP) \cite{bernstein2005bounded}. A Dec-POMDP is a tuple $(I, \mathcal{S}, \{\mathcal{A}_i\}, \{\mathcal{O}_i\}, P, R)$ where $I$ is the finite set of agents indexed $1$ to $M$, $S$ is the set of states, $\mathcal{A}_i$ is the set of actions for agent $i$, and thus $\prod_{i=1}^M \mathcal{A}_i$ is the joint action space, $\mathcal{O}_i$ is the observation space of agent $i$, and thus $\prod_{i=1}^M \mathcal{O}_i$ is the joint observation space, $P = P(\textbf{s}', \textbf{o} | \textbf{s}, \textbf{a})$ (where $\textbf{o} = (o_1,\ldots,o_M)$ and similarly for the others) is the state-transition probability for the whole system, and $R:\mathcal{S} \times \prod_{i=1}^M\mathcal{A}_i \rightarrow \mathcal{R}$ is the reward. 

In the case when the joint observation $\textbf{o}$ equals the world state of the system, then we call the system a decentralized Markov decision process (Dec-MDP).

\textbf{DAgger:} The Dataset Aggregation (DAgger) algorithm \cite{DAgger} is an iterative imitation learning algorithm that seeks to learn a policy from expert demonstration. The main idea is to allow the learning policy to navigate its way through the environment, and have it query the expert on states that it sees. It does this by starting with a policy $\hat{\pi}_2$ which learns from the dataset of expert trajectories $\mathcal{D}_1$ through supervised learning. Using $\hat{\pi}_2$, a new dataset is generated by rolling out the policy and having the expert provide supervision on the decisions that the policy made. This new dataset is aggregated with the existing set into $\mathcal{D}_2 \supset \mathcal{D}_1$. This process is iterated, i.e.\ a new $\hat{\pi}_3$ is trained, another new dataset is obtained and aggregated into $\mathcal{D}_3 \supset \mathcal{D}_2$ and so on. Learning in this way has been shown to be more stable and have nicer convergence properties as learning utilizes trajectories seen from the learner's state distribution, as opposed to only the expert's state distribution.

\textbf{Policy Gradients (PG):} One approach to RL problems are policy gradient methods \cite{NIPS1999_1713}: instead of directly learning state-action values, the parameters $\theta$ of the policy $\pi_\theta$ are adjusted to maximize the objective,
\begin{equation*}
J(\theta) = \mathbb{E}_{s\sim p^\pi, a\sim \pi_\theta}\left[Q^\pi(s,a) \right], 
\end{equation*}
where $p^\pi$ is the state distribution from following policy $\pi$. The gradient of the above expression can be written as \cite{NIPS1999_1713, SuttonB18}:
\begin{equation*}
\nabla_\theta J(\theta) = \mathbb{E}_{s\sim p^\pi, a\sim \pi_\theta} [(\nabla_\theta \log \pi_\theta(s|a)) Q^\pi(s,a)]. 
\end{equation*}
Many policy gradient methods seek to reduce the variance of the above gradient estimate, and thus study how one estimates $Q^\pi(s,a)$ above \cite{SchulmanMLJA15}. For example, if we let $Q^\pi(s,a)$ be the sample return $R^t = \sum_{i=t}^T \gamma^{i-t} r_i$, then we get the REINFORCE algorithm \cite{REINFORCE}. Or one can choose to learn $Q^\pi(s,a)$ using temporal-difference learning \cite{Sutton1988LearningTP, SuttonB18}, and would obtain the Actor-Critic algorithms \cite{SuttonB18}. Other policy gradients algorithms are: DPG \cite{SilverDPG}, DDPG \cite{LillicrapHPHETS15}, A2C and A3C \cite{MnihBMGLHSK16}, to name a few.

Policy Gradients have been applied to multi-agent problems; in particular the Multi-Agent Deep Deterministic Policy Gradient (MADDPG) \cite{maddpg} uses an actor-critic approach to MARL, and this is the main baseline we test our method against. Another policy gradient method is by \cite{FoersterFANW17} called Counterfactual Multi-Agent (COMA), who also uses an actor-critic approach.

\section{Methods}

In this section, we explain the motivation and method of our approach: Centralized Expert Supervises Multi-Agents (CESMA), which falls under the popular framework of centralized training but decentralized execution.

\subsection{Treating a multi-agent problem as a single-agent problem}\label{subsec:matosa}

Intuitively, an optimal strategy of a multi-agent problem could be found by a centralized expert with full observability. This is because the centralized controller has the most information available about the environment, and therefore would not pay a high of cost of partial-observability that independent learners might. 

To find this centralized expert, we treat a multi-agent problem as a single agent problem in the joint observation and action space of all agents. This is done by concatenating the observations of all agents into one observation vector for the centralized expert, and the expert learns outputs that represent the joint actions of the agents.

Our framework does not impose any other particular constraints on the expert. Any expert architecture that outputs an action that represents the joint-actions of all of the agents may be used.  Due to that, we are free to use any standard RL algorithm for the expert such as DDPG, DQN, or potentially even analytically derived experts.

\edit{
In the case of collaborative multi-agent reinforcement learning, the multi-agents have a shared reward so that there is global objective function that must be maximized based on the collaborative efforts of the agents:}
    \begin{equation*}
        \edit{\max_{(\pi_1,\ldots,\pi_M)} V_{(\pi_1,\ldots,\pi_M)}(s_0) = \mathbb{E}\left[ \sum_{t=0}^{T-1} \gamma^t R(\textbf{a}_t, s_t) \large\mid s_0, (\pi_1\ldots, \pi_M) \right]}
    \end{equation*}
    \edit{where $\gamma$ is a discount factor (moreso needed in the case that $T=\infty$), $s_t$ represents a global state vector, and $\textbf{a}_t = (a_1,\ldots, a_M) = (\pi_1(o_1),\ldots, \pi_M(o_M))$, where there is an observation function $f(s_t) = (o_1,\ldots, o_M)$ which outputs the observations of the individual agents.}
    
    \edit{We re-emphasize that the goal of Multi-Agent Reinforcement Learning is to obtain agents that only take in as input their local observations during the execution phase.}

\subsection{Curse of dimensionality and some reliefs}\label{subsec:cursedim}

When training a centralized expert, both the observation space and action space can grow exponentially. For example, if we use a DQN for our centralized expert then the number of output nodes will typically grow exponentially with respect to the number of agents. This is due to each output needing to correspond to an element in the joint action space $\prod_{i=1}^M \mathcal{A}_i$.

One way to deal with the exponential growth in the joint action space is, rather than requiring the centralized expert to move all agents simultaneously, we can restrict it to moving only one agent at a time, while the others default to a ``do nothing" action (assuming one is available). Effectively this mean the growth in the action space is now linear with respect to the number of agents. We provide an experiment where we were able to decentralize such an expert in Section~\ref{subsec:dqns-otex}.

This problem has also been studied by QMIX \cite{qmix} and VDNs (Value Decomposition Networks) \cite{sunehag2017value}, where exponential scaling of the output space is solved by having separate $Q$ values for each agent and then using the sum as a system $Q$. Due to the nature of the reduction technique, these approaches require their own theorems of convergence. Other techniques such as action branching \cite{tavakoli2018action} have been considered. An experiment where we decentralize QMIX/VDN-like centralized expert models (which grow linearly in the number of output nodes) can be found in Section Section~\ref{subsec:dqns-otex}. 

In our experiments, we use DDPG (with Gumbel-Softmax action selection if the environment is discrete, as MADDPG does also) to avoid the exploding number of input nodes of the observation space, as well as exploding number of output nodes of the action space. Under this paradigm, the input and output nodes only grow linearly with the number of agents, as the output nodes of a neural network in DDPG is the chosen joint action, as opposed to a DQN, where the output nodes must enumerate all possible joint actions.

\subsection{CESMA for multi-agents without communication}\label{subsec:cesmanoncom}

\begin{figure}
	\centering
	\includegraphics[scale=0.3]{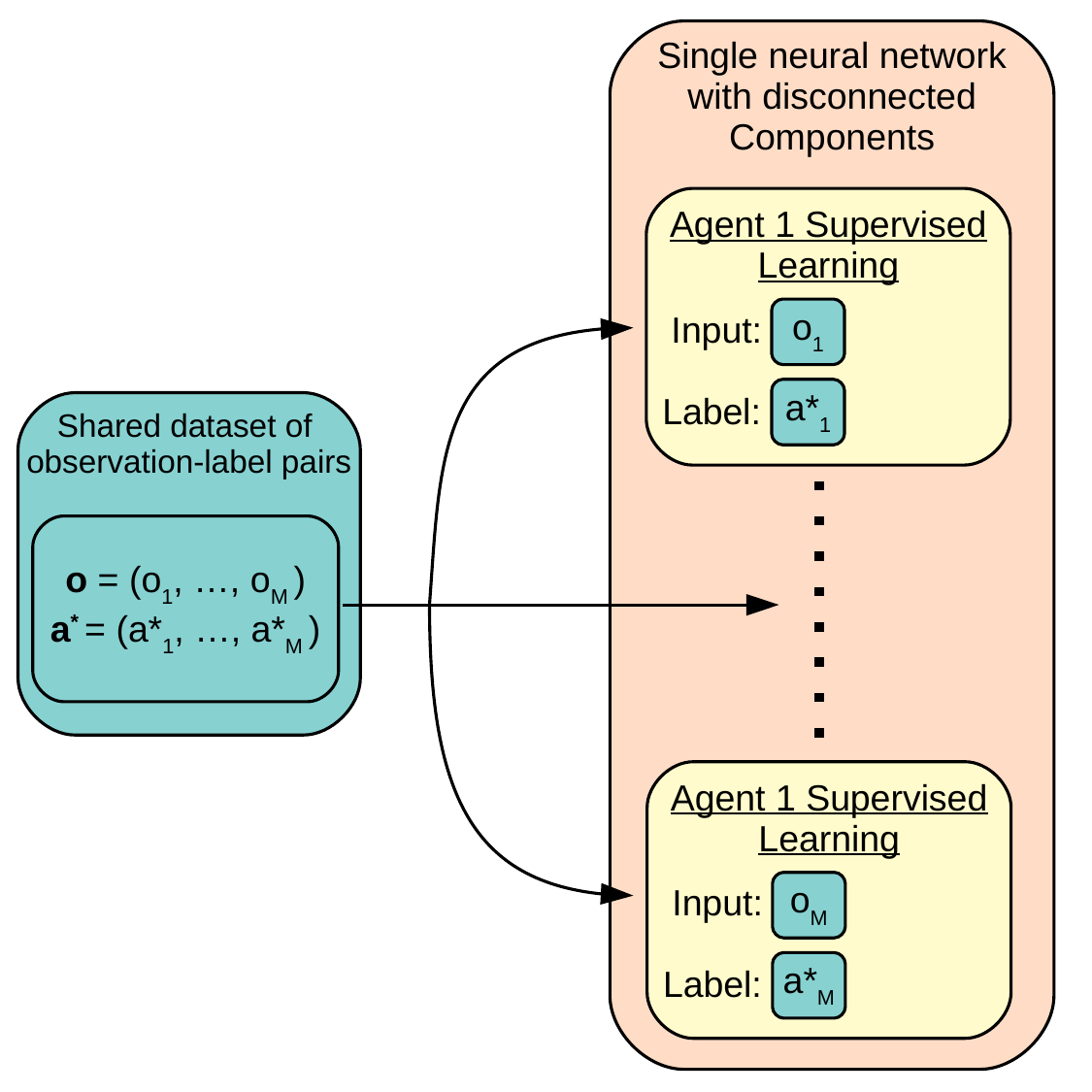}
	\caption{The centralized expert labels guide supervised learning for the multi-agents. The multi-agents make up the disconnected components of a single-agent learner.}
	\label{fig:supervise_multiagents}
\end{figure}

To perform imitation learning to decentralize the expert policy, we adapt DAgger to the multi-agent setting. But we note that the proposed framework could handle a myriad of imitation learning algorithms, such as Forward Training \cite{ross2010efficient}, SMILe \cite{ross2010efficient}, SEARN \cite{daume2009search}, and more.

There are many ways DAgger can be applied to multi-agents, but we implement a method that best allows the theoretical analysis from \cite{DAgger} to apply: Namely after training the expert, we do supervised learning on a single neural network with disconnected components, each corresponding to one of the agents. 

In more detail, after training a centralized expert $\pi^*$, we initialize the $M$ agents $\pi_{1}, \ldots, \pi_{M}$, and initialize the dataset of observation-label pairs $\mathcal{D}$. The agents then step through the environment, storing each observation $\textbf{o} = (o_1,\ldots, o_M)$ (where $o_i$ is agent $i$'s observation) the multi-agents encounter, along with the expert action label ${\textbf{a}}^*=\pi^*(\textbf{o})$ (where ${\textbf{a}}^* = ({a}_1^*,\ldots, {a}_M^*)$ and ${a}_i^*$ is agent $i$'s expert label action); so we store the pair $(\textbf{o}, {\textbf{a}}^*)$ in $\mathcal{D}$ at each timestep. After $\mathcal{D}$ has reached a sufficient size, at every $k$th time step (chosen by the practitioner; we used $k=1$ in our experiments), we sample a batch from this dataset $\{(\textbf{o}^{(\beta)}, {\textbf{a}}^{*,(\beta)})\}_{\beta=1}^B$, and then distribute the data batch $\{(o^{(\beta)}_i,{a}^{*,(\beta)}_{i})\}_{\beta=1}^B$ to agent $i$, for supervised learning; we note the training can be done sequentially or parallel. Having a shared dataset of trajectories in this way allows us to view $(\pi_{1}, \ldots, \pi_{M})$ as a single neural-network with disconnected components, and thus the error bounds from \cite{DAgger} directly apply, as discussed in Section \ref{sec:theory}. See Figure \ref{fig:supervise_multiagents} for a diagram. Pseudo-code for our method is contained in Appendix \ref{sec:pseudocodeCESMAwithout}. (In Appendix \ref{subsec:individual_dataset} we test whether giving each agent its own dataset would make a difference, and it did not seem so).

The aforementioned procedure is sufficient when the agents do not need to communicate, but when communication is involved we have to modify the above method.

\subsection{CESMA for multi-agents with communication}\label{subsec:cesma_comm}

The main insight for training an agent's communication action is that we can view a broadcasting agent and the receiving agent as one neural network connected via the communication nodes; then in this way we can backpropagate the action loss of the receiving agent through to the broadcasting agent's weights. 

\begin{figure}[t]
	\centering
	\begin{minipage}{0.7\textwidth}
		\centering
		\includegraphics[width=\textwidth]{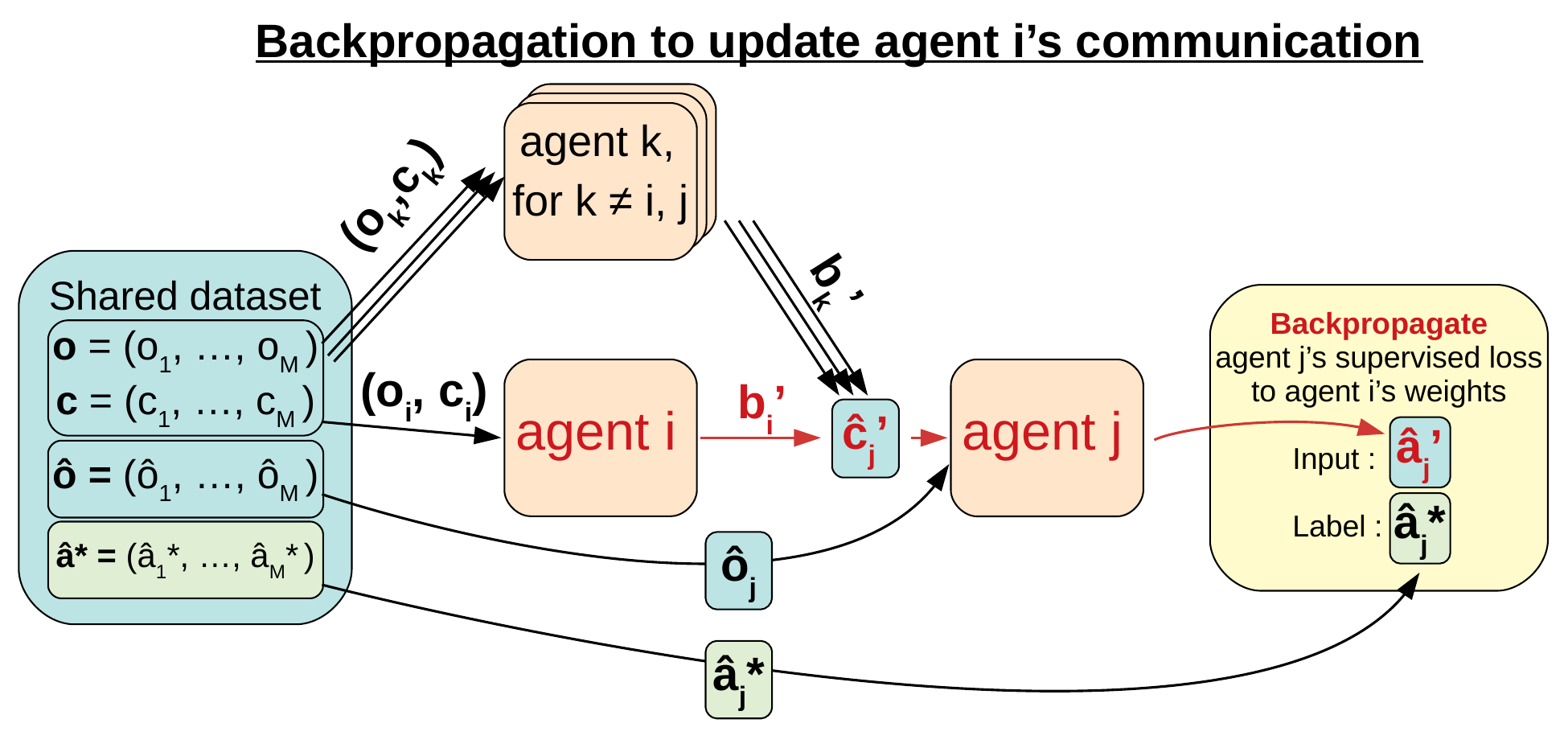}
	\end{minipage}
	\begin{minipage}{0.7\textwidth}
		\centering
		\includegraphics[width=\textwidth]{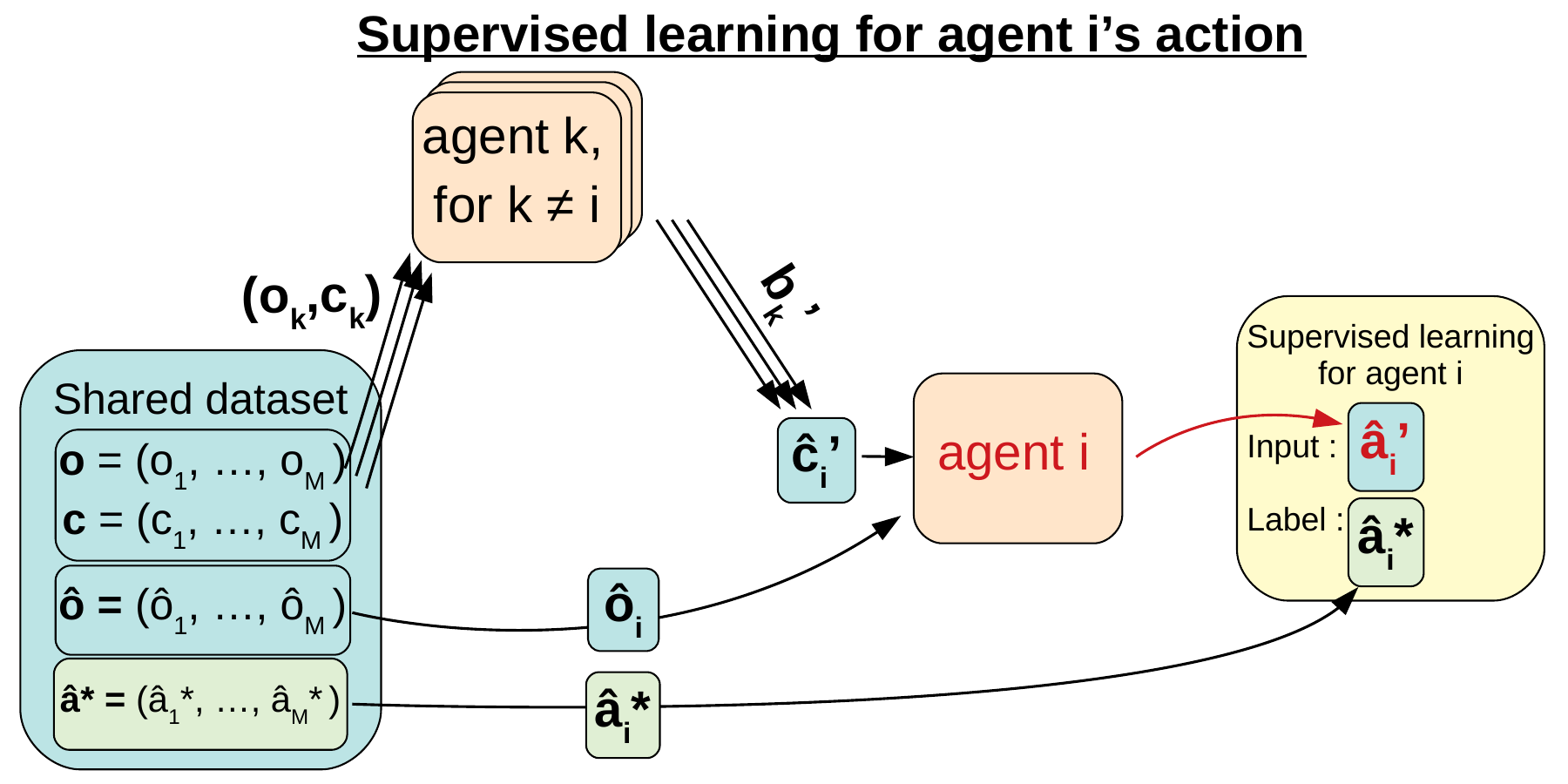}
	\end{minipage}
	\caption{Decentralizing multi-agents that communicate. The top diagram shows how we update agent $i$'s communication action by backpropagating the supervised loss of \emph{other} agents. The red portions highlight the trail of backpropagation. The bottom diagram shows how we update the action of agent $i$.}
	\label{fig:cesma_with_comm}
\end{figure}

In more detail, due to communication, the multi-agents now have two types of observations and actions. Thus, we denote the physical  actions (i.e.\ non-communication actions) as $\textbf{a} = (a_1, \ldots, a_M)$ and the communication actions/\textbf{b}roadcasts as $\textbf{b} = (b_1, \ldots, b_M)$. \emph{For notational simplicity, let us assume that all agents can communicate with each other and each agent broadcasts the same thing to all other agents.} So we denote $c_i = (b_1,\ldots, b_{i-1}, b_{i+1}, \ldots, b_M)$ as agent $i$'s observation of the broadcast by other agents, and where $b_j$ is agent $j$'s broadcast to all other agents. So for each agent $i$, we have $\pi_i(o_i, c_i) = (a_i, b_i)$. And we also denote $\pi_i(o_i, c_i)_{\text{action}} = a_i$, and $\pi_i(o_i, c_i)_{\text{comm}} = b_i$. 

For training, as before we have a shared dataset of observations $\mathcal{D}$. But as the agents step through the environment, at each timestep we now store $((\textbf{o}, \textbf{c}), \hat{\textbf{o}}, \hat{\textbf{a}}^*)$, where $(\textbf{o}, \textbf{c})$ is the joint physical and communication observation of the previous timestep, $\hat{\textbf{o}}$ is the physical observation at the current timestep, and $\hat{\textbf{a}}^* = \pi^*(\hat{\textbf{o}}) = (\hat{a}_1^*,\ldots, \hat{a}_M^*)$ is the expert action label; these are the necessary ingredients for training.

Then to train, we first obtain a sample from $\mathcal{D}$ (practically we perform batched training, but for simplicity we consider one sample), say $((\textbf{o}, \textbf{c}), \hat{\textbf{o}}, \hat{\textbf{a}}^*)$, and then we take the policies at the most-recent update $\pi_1^{\text{current}},\ldots, \pi_M^{\text{current}}$ and form their broadcasts $b_k' = \pi^{\text{current}}_k(o_k, c_k)_{\text{comm}}$ for $k=1,\ldots, M$. Then in principle, we want to minimize the loss function,
\begin{equation*}
\min_{(\pi_1,\ldots, \pi_M)} \sum_{j=1}^M \ell(\hat{a}^*_j, \;\pi_j(\hat{o}_j, \hat{c}_j)_{\text{action}}), 
\end{equation*}
where
\begin{equation*}
\hat{c}_j = (b_1' , \ldots, b_{j-1}', b_{j+1}', \ldots, b_M'), \quad j=1,\ldots, M.
\end{equation*}
In practice, we train each agent $i$ separately by minimizing their \emph{communication} loss and \emph{action} loss which we describe below. 

In order to train agent $i$'s communication action, we make the insight that we can backpropagate the supervised learning loss of \emph{other} agents through the communication nodes to agent $i$'s parameters, precisely because the communication \emph{output} of agent $i$ becomes an observational \emph{input} for the \emph{other} agents. 
Then to train the communication action of agent $i$, we sample $((\textbf{o}, \textbf{c}), \hat{\textbf{o}}, \hat{\textbf{a}}^*)$ from $\mathcal{D}$, and seek to minimize the communication loss function,
\begin{equation*}
\min_{\pi_i} \sum_{
	j\neq i}
\ell(\hat{a}^*_j, \; \pi_j(\hat{o}_j, \hat{c}_j')_{\text{action}}) \tag{comm.\ loss for agent $i$}, 
\end{equation*}
where 
\begin{equation*}
\hat{c}_j' = (b_{1}', \ldots, \pi_i(o_i, c_i)_{\text{comm}}, \ldots, b_{j-1}', b_{j+1}', \ldots, b_{M}'), 
\end{equation*} 
where we assumed without loss of generality that $i< j$. And so because $\hat{c}_j'$ %'
\emph{depends on} $\pi_i$, then we can backpropagate agent $j$'s supervised loss to agent $i$'s parameters.
To train the physical action of agent $i$, we sample $((\textbf{o}, \textbf{c}), \hat{\textbf{o}}, \hat{\textbf{a}}^*)$ from $\mathcal{D}$ and want to minimize
\begin{equation*}
\min_{\pi_i} \ell(\hat{a}_i^*, \; \pi_i(\hat{o}_i, \hat{c}_i')_{\text{action}} \tag{action loss for agent $i$}),  
\end{equation*}
where $\hat{c}_i' = (b_{1}',\ldots, b_{i-1}', b_{i+1}', \ldots, b_{M}')$.

For a graphic overview, we give a diagram in Figure \ref{fig:cesma_with_comm} for the backpropagation of the communication loss and the action loss, and provide pseudocode in Algorithm~\ref{alg:supervising_multiagents_comm} in Appendix~\ref{sec:pseudocodeCESMAwithout}. In some sense, our method can be viewed as a hybrid of experience replay and supervised learning.

In this way, we have alleviated a bit the issue of sparse rewards for communication \citep[][Section 4]{learntocommunicate}. % (Section 4).
Indeed, communication actions suffer from sparse rewards as a reward is only bestowed on the broadcasting agent when all the following align: it sends the right message, the receiving agent understands the message, and then acts accordingly. In our method with an expert supervisor, the correct action by the acting agent is clear.

\section{Theoretical analysis: No-regret analysis and guarantees}\label{sec:theory}

In our approach we are adapting \cite{DAgger} to the multi-agent setting, and thus we present a direct rephrasing of \citep[][Theorem~3.2]{DAgger}. This is possible because we can view the multi-agents as a single-agent learner with disconnected components (as described in Section \ref{subsec:cesmanoncom}). This analysis takes the form of a no-regret analysis, and so provides theoretical guarantees on the reward obtainable by the agents (which may not be the same as the expert). Notationally, 
\begin{itemize}
	\item we let $\ell$ be a surrogate loss of matching the expert policy $\pi^*$ (e.g.\ the expected 0-1 loss at each state) and denote $r=r(s,a)$ the instantaneous reward which we assume to be bounded in $[0,1]$, 
	\item $(\pi^{(N)}_1, \ldots, \pi^{(N)}_M)$ are the multi-agents after $N$ updates of the policy using any supervised learning algorithm, and where each update is done after a $T$-step trajectory with $T$ the task horizon, 
	\item $d_{(\pi^{(N)}_1, \ldots, \pi^{(N)}_M)}$ is the average distribution of observations that come from following the multi-agent policy $(\pi^{(N)}_1, \ldots, \pi^{(N)}_M)$ from a given initial distribution, 
	\item $R(\pi^{(N)}_1, \ldots, \pi^{(N)}_M)$ is the cumulative reward after an episode of the task,
	\item and $U^{\pi'}_t(s,\pi)$ is the reward after $t$ steps of executing $\pi$ in only initial state $s$, and then following policy $\pi'$ thereafter.
\end{itemize}
Then viewing the multi-agent policy as a joint single-agent policy we obtain the following \emph{guarantee on the reward based on how well the multi-agents match the expert}:
\begin{theorem}\label{thm:TCP}
	If the number of policy updates $N$ is $O(T\log^k(T))$ for sufficiently large $k\ge 0$, then there exists a joint multi-agent policy $(\hat{\pi}_1, \ldots, \hat{\pi}_M) \in \{({\pi}^{(i)}_1, \ldots, {\pi}^{(i)}_M)\}_{i=1}^N$ such that
	\begin{equation*}
	R(\hat{\pi}_1, \ldots, \hat{\pi}_M) \ge R(\pi^*) - uT\mu_N - O(1),  
	\end{equation*}
	where $u\ge 0$ is such that $U^{\pi^*}_{T-t+1}(s, \pi^*) - U^{\pi^*}_{T-t+1}(s,a) \leq u$ for all actions $a$ and $t\in \{1,\ldots, T\}$, and
	\begin{align*}
	\mu_N = \min_{(\pi_1, \ldots, \pi_M)} &\frac{1}{N} \sum_{i=1}^N \mathbb{E}_{\textbf{o}\,\sim\, d_{ (\pi^{(i)}_1, \ldots, \pi^{(i)}_M) }} [\ell(\textbf{o}, (\pi_1, \ldots, \pi_M)]. 
	\end{align*}
\end{theorem}
%The proof can be found in \cite{DAgger}.
Here $\mu_N$ is best described as the true loss of the best learned policy in hindsight. The condition $U^{\pi^*}_{T-t+1}(s, \pi^*)-U^{\pi^*}_{T-t+1}(s,a) \leq u$ can best be described as saying the reward lost from not following the expert at initial state $s$, but following it after, is at most $u$. We further remark that $\mu_N$ is a bound on the performance gap between the multi-agents and centralized expert that may not necessarily vanish, but rather is best viewed as a guarantee on the reward obtainable by the multi-agents. The role of partial observability and communication can be considered orthogonal to the current discussion so we leave this in the Appendix (Section \ref{sec:partobscomm}). \ATLNote{(This pointer seems okay, can you make it cleaner?)}

\ATLNote{(Maybe add more discussion about the above theorem?)}

\section{Experiments}

\subsection{Comparison with Decentralized Learning in a Complex Environment}

\begin{figure}[h]
    \centering
    \includegraphics[width=0.6\textwidth]{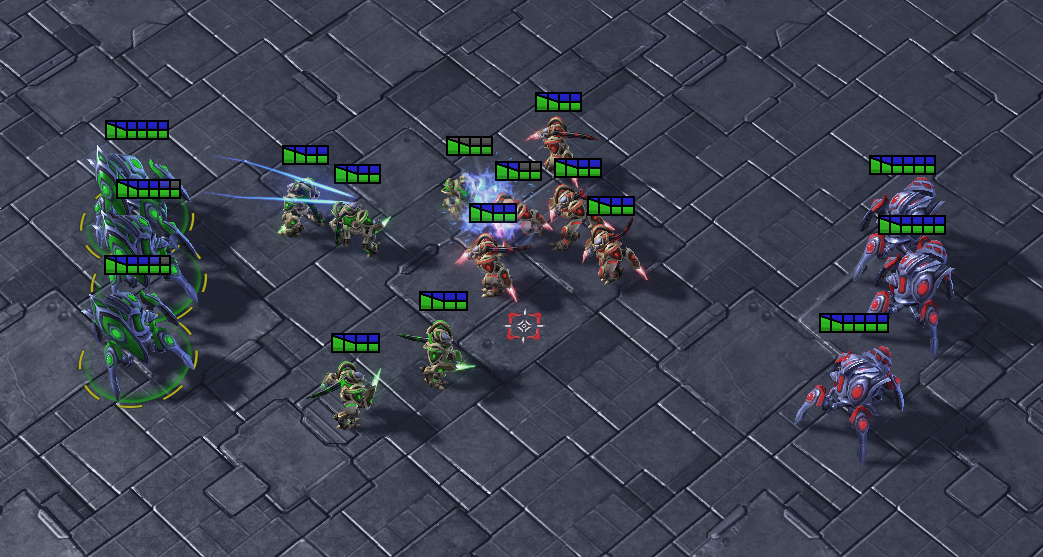}
    \caption{A screenshot of the StarCraft 2 environment used in our experiment. The green-colored units are the allies, and the red-colored units are the enemies. This is an 8 vs 8 environment against an AI coded by the game developers, under the ``very hard'' difficulty. The units are heterogeneous, and due to sight-range/limited visibility, each agent only has partial observability (i.e. fog of war).}
    \label{fig:sc2}
\end{figure}

\edit{
Here we compare our centralized learning (but decentralized execution) method with decentralized learning. In order to highlight the capabilities of CESMA, as well as the pitfalls of decentralized learning, we conduct an experiment in a complex environment developed for multi-agent training in the popular strategy videogame, StarCraft 2 \cite{samvelyan2019starcraft}. In this environment, each unit receives the following local observations for each allied and enemy unit: distance, relative $x$, relative $y$, health, shield, and unit type. The \emph{discrete} action space consists of: movement (4 directions), an attack action available for each enemy unit, a stop, and a no-op (dead agents can only take no-op). The overall goal is to maximize the win rate, but rewards are given based on hit-point damage dealt and enemy units killed, as well as a special bonus for winning the battle. The units also have sight-range/limited visibility, so this is a partially observable environment from the perspective of each agent. Here, our multi-agents play an 8 vs 8 battle with heterogeneous units (3 Stalker units and 5 Zealots units), against AI created by game developers under the ``very hard" difficulty (the hardest one before allowing enemies to cheat). See Figure \ref{fig:sc2} for a screenshot of the environment.
}

\edit{
In our experiments, we follow the comparison procedures in \cite{samvelyan2019starcraft}, and use their implementation of the decentralized learning method, Independent Q-Learning (IQL). Thus our multi-agents consists of recurrent neural networks (RNNs) with 64 hidden units, and we use five independent runs of IQL and CESMA to evaluate performance -- namely we measure the test battle-win percentage. Our expert neural network has the same architecture as the multi-agents, but now receives the concatenated global observations as input, and outputs Q-values for each agent, and we take the sum of these Q-values as the expert Q-value -- this means the action space only grows linearly with the multi-agents, just like in the IQL case. This is well justified in \cite{sunehag2017value, qmix}. After training the centralized expert, we decentralize the one with the highest test win percentage in order to obtain decentralized multi-agents. Further experimental setup details can be found in the Appendix \ref{app:sc2}, and hyperparameters in \ref{app:sc2_hyp}.
}

\begin{figure}
    \centering
    \includegraphics[width=0.8\textwidth]{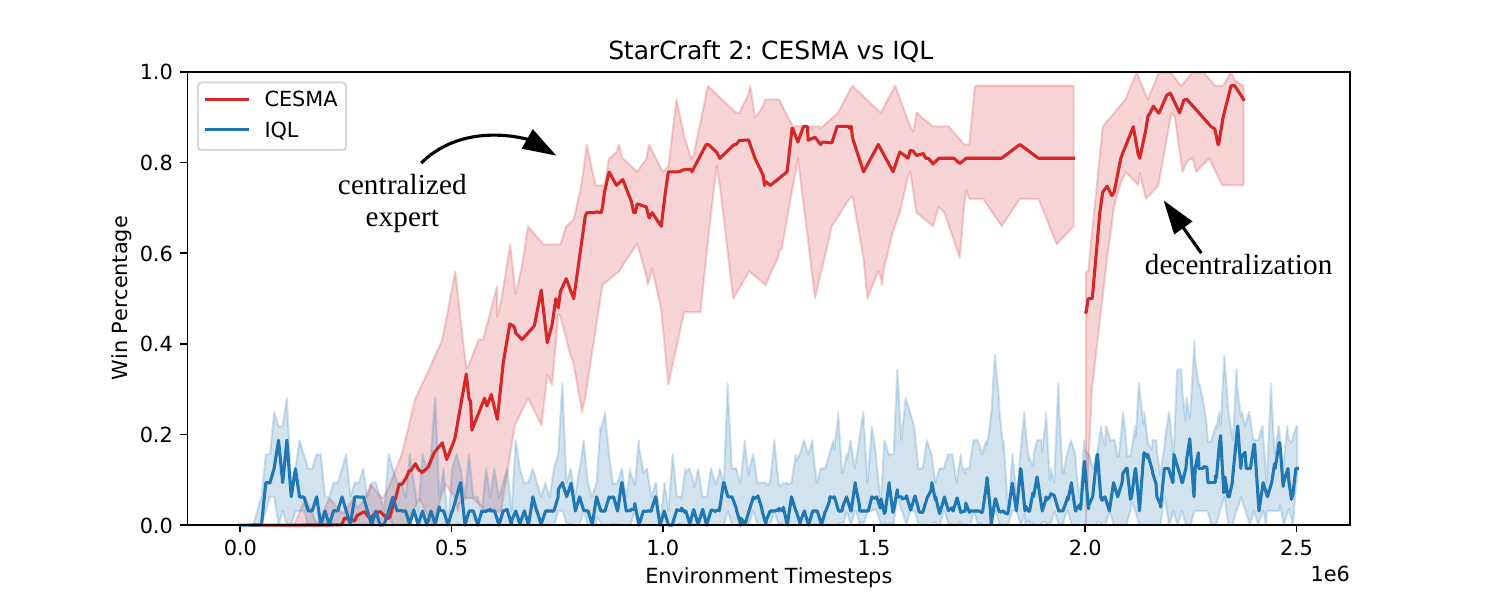}
    \caption{Learning curves for the StarCraft 2 multi-agent environment based on five independent runs. The middle bold line represents the median win percentage, and the envelopes represent the max and min. We compare our method (CESMA) with a decentralized learning method (IQL). For CESMA, the first red curve represents training of the centralized expert, and the second red curve represents decentralization. As can be seen, this environment is tough for a decentralize learner, but utilizing centralized training and then decentralization CESMA is able to achieve high win-rates (sometimes even 100\% win rates).}
    \label{fig:sc2_learning_curves}
\end{figure}

\edit{
The results can be seen in Figure \ref{fig:sc2_learning_curves}, where we plot the median win percentage, as well as the max and min envelopes, over five independent runs. As can be seen, this environment is tough for the decentralized learner, but for our centralized expert we are able to achieve high win-rates, which can then be decentralized to find the same high win-rates (sometimes even 100\%). This shows we can effectively decentralize even complex centralized experts in complicated environments such as StarCraft 2.
}

\edit{
Later in Section \ref{subsec:decent_learning_comm}, we discuss the difficulty that decentralized learners have in learning how to communicate.
}

\subsection{Cooperative Navigation}\label{subsec:coopnav}

Here, our experiments are conducted in the Multi-Agent Particle Environment \cite{mordatch2017emergence, maddpg} provided by OpenAI. In order to conduct comparisons to MADDPG, we also use the DDPG algorithm with the Gumbel-Softmax \cite{jang2016categorical, maddison2016concrete} action selection as they do. For the single-agent centralized expert neural network, we always make sure the number of parameters/weights matches (or is lower) than that of MADDPG's. For the decentralized agents, we use the same number of parameters as the decentralized agents in MADDPG (i.e.\ the actor part). Following their experimental procedure, we average our experiments over three runs, and plot the minimum and maximum reward envelopes. And for the decentralization, we trained three separate centralized experts, and used each of them to obtain three decentralized policies. Full details of our hyperparameters and the environments are in the appendix. \edit{We note MADDPG cannot be used in the above environment, StarCraft 2, as it only works on continuous action spaces.}

Here we examine the situation of $N$ agents occupying $N$ landmarks in a 2D plane, and the agents are either homogeneous or nonhomogenous, and have control over their acceleration. \edit{They are also allowed to collide and bounce off each other, although collisions are penalized in this environment, and thus to achieve a high reward collisions must be avoided.} The (continuous) observations of each agent are the relative positions of other agents, the relative positions of each landmark, and its own velocity. The agents do not have access to others' velocities so we have partial observability. The reward is based on how close each landmark has an agent near it, and the actions of each agent are discrete: up, down, left, right, and do nothing.

In Figure \ref{fig:learning_curves}, we see that CESMA, when combining the number of samples in training the expert as well as decentralization, is able to achieve the same reward as MADDPG while utilizing fewer samples, i.e.\ CESMA is more sample efficient (the dashed red line is just a visual aid that extrapolates the reward for the decentralized curves, because we stop training once the reward sufficiently matches MADDPG). 

In Figure \ref{fig:better_opt}, we also noticed that the centralized expert is able to find a policy that achieves a higher reward than a \emph{converged} MADDPG; and we were able to decentralize this expert to obtain decentralized multi-agent policies that achieved higher rewards than MADDPG.

\newcommand{\figscale}{0.32}

\begin{figure}[t]
	\begin{minipage}{\figscale\linewidth}
		\centering
		\includegraphics[width=\linewidth]{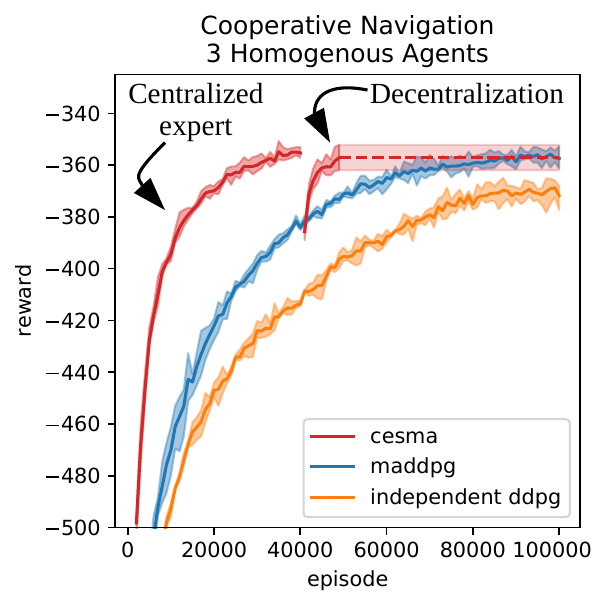}
		%		\label{fig:simple_spread}
	\end{minipage}%
	\begin{minipage}{\figscale\linewidth}
		\centering
		\includegraphics[width=\linewidth]{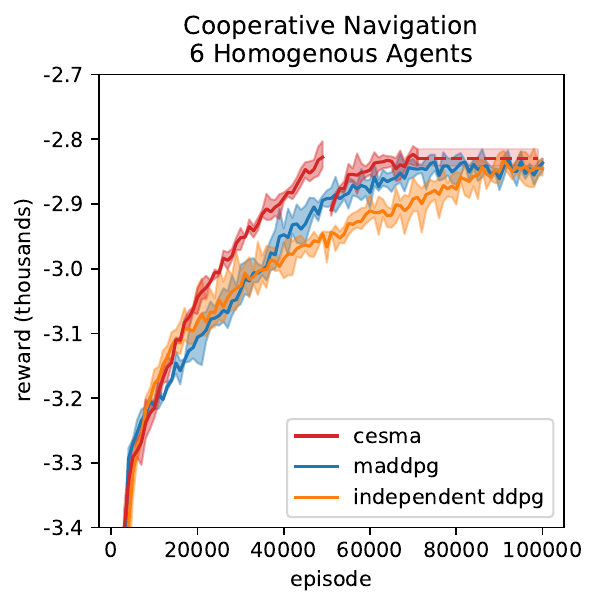}
		%		\label{fig:simple_spread_many6_1bound}
	\end{minipage}%
	\begin{minipage}{\figscale\linewidth}
		\centering
		\includegraphics[width=\linewidth]{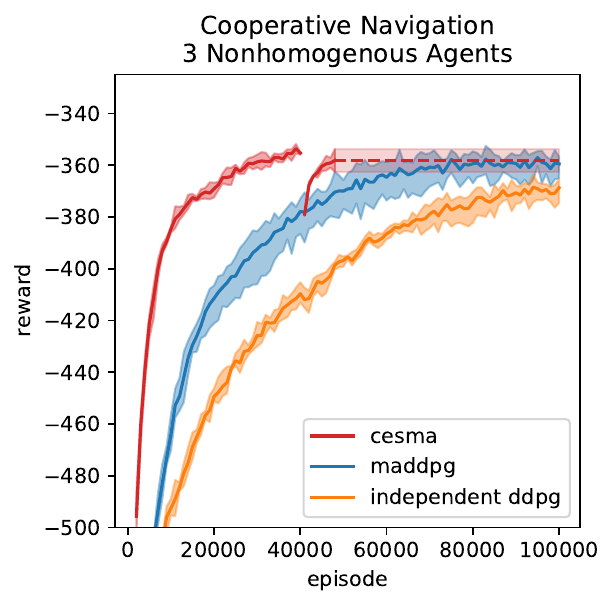}
	\end{minipage}
	\begin{minipage}{\figscale\linewidth}
		\centering
		\includegraphics[width=\linewidth]{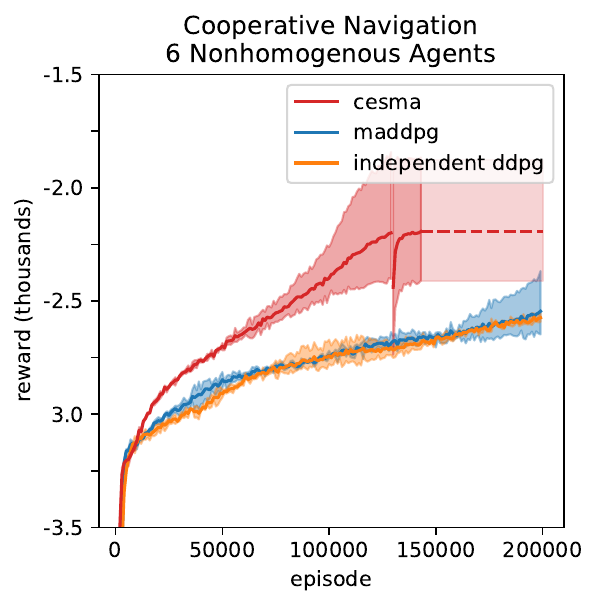}
	\end{minipage}%
	\begin{minipage}{\figscale\linewidth}
		\centering
		\includegraphics[width=\linewidth]{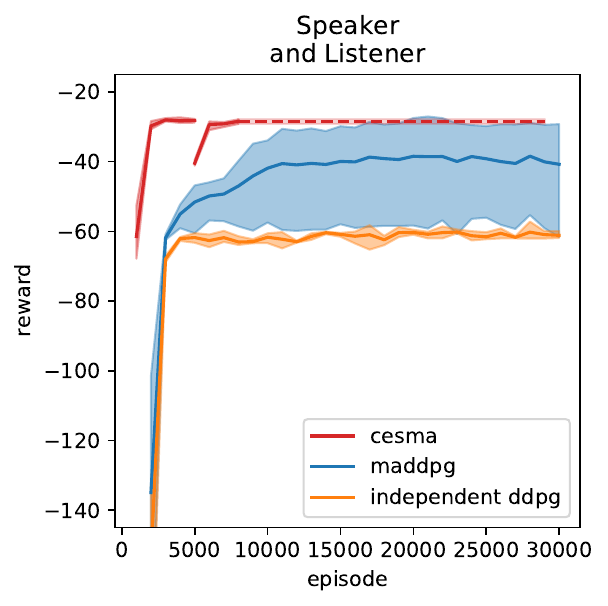}
	\end{minipage}%
	\begin{minipage}{\figscale\linewidth}
		\centering
		\includegraphics[width=\linewidth]{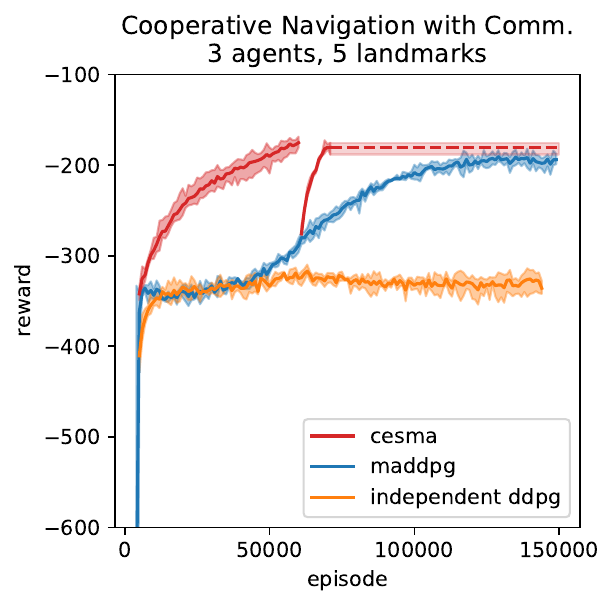}
	\end{minipage}
	\caption{Reward curves for various multi-agent environments. We train the centralized expert until its reward matches or betters MADDPG's reward. Then we decentralize this expert until we achieve the same reward as the expert. The first red curve is the reward curve of the centralized expert, and the second red curve is reward curve for the decentralized agents. The dashed red line is a visual aid extrapolating the reward for the decentralized curves, because we stop training the agents once the reward matches the expert. We see that CESMA is more sample-efficient than MADDPG.}
	\label{fig:learning_curves}
\end{figure}%
\begin{figure}[t]
	\begin{minipage}{\figscale\textwidth}
		\centering
		\includegraphics[width=\textwidth]{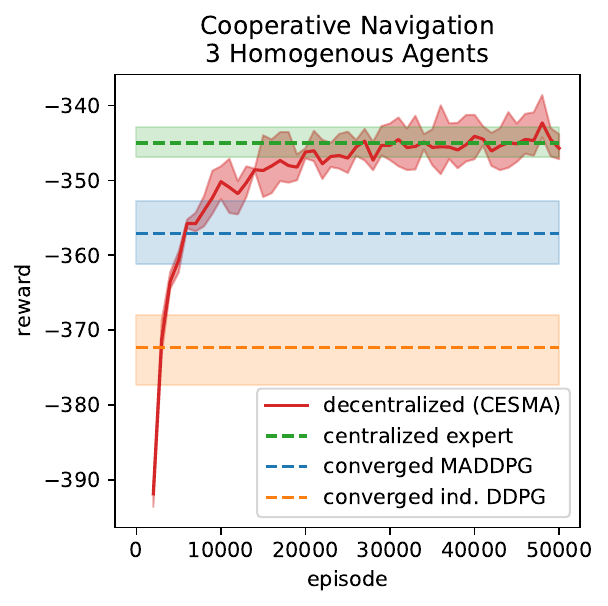}
	\end{minipage}%
	\begin{minipage}{\figscale\linewidth}
		\centering
		\includegraphics[width=\linewidth]{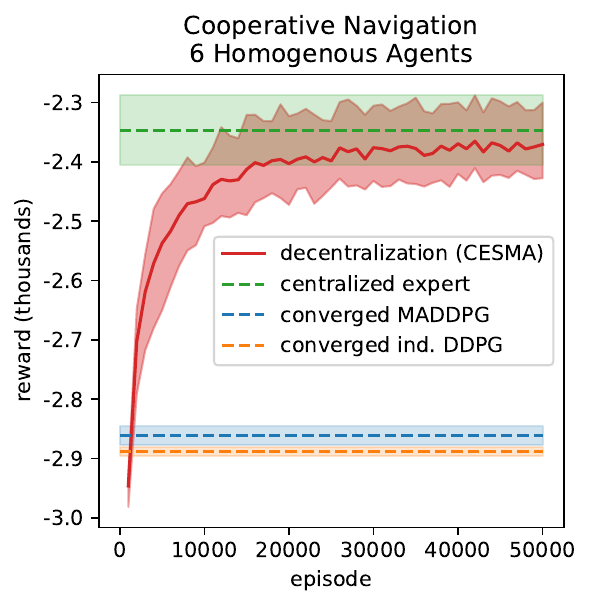}
	\end{minipage}%
	\begin{minipage}{\figscale\textwidth}
		\centering
		\includegraphics[width=\textwidth]{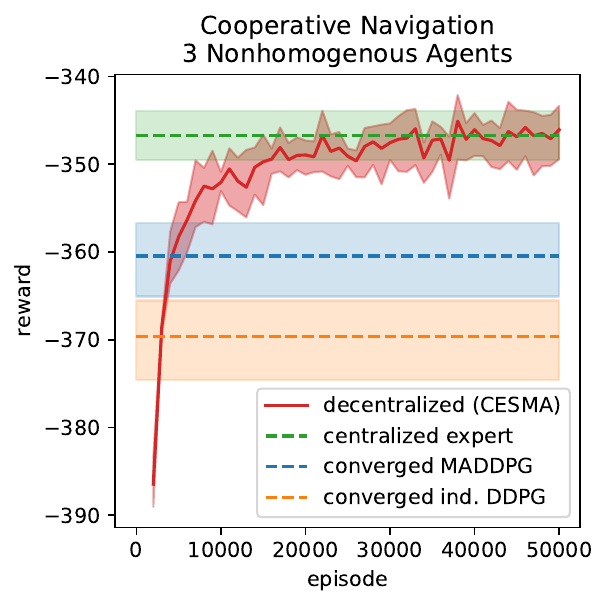}
	\end{minipage}
	\begin{minipage}{\figscale\textwidth}
		\centering
		\includegraphics[width=\textwidth]{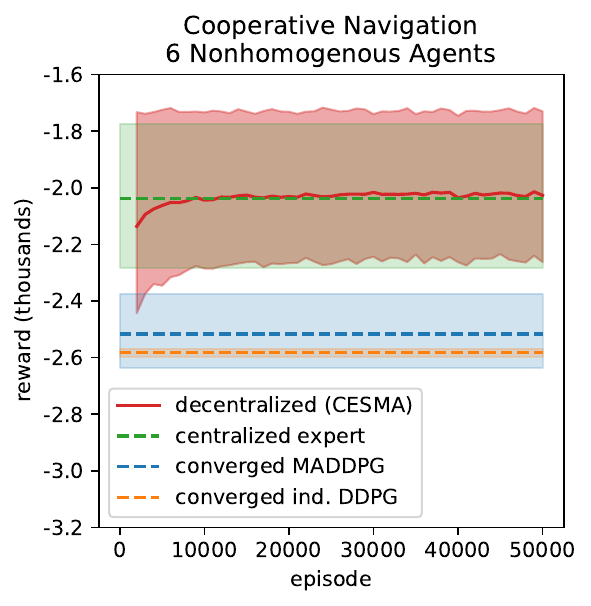}
	\end{minipage}%
	\begin{minipage}{\figscale\linewidth}
		\centering
		\includegraphics[width=\linewidth]{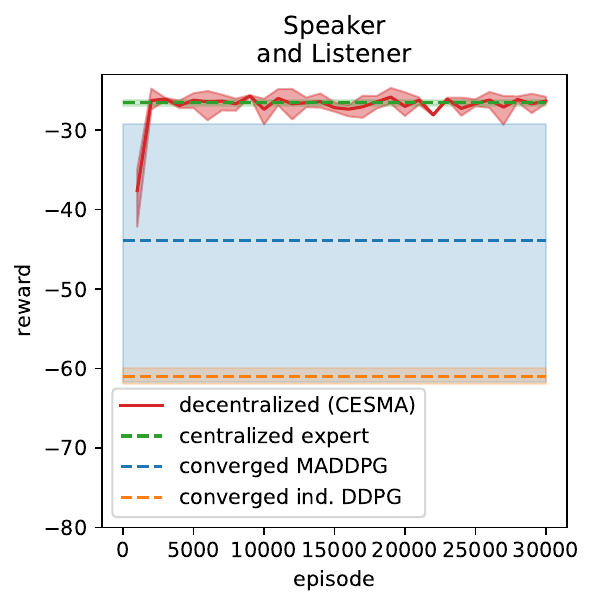}
	\end{minipage}%
	\begin{minipage}{\figscale\linewidth}
		\centering
		\includegraphics[width=\linewidth]{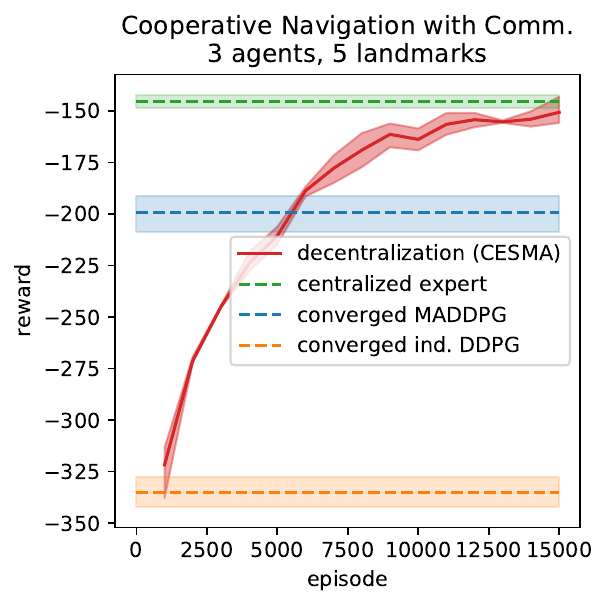}
	\end{minipage}%
	\caption{Reward curves for decentralization of a centralized expert policy that obtains a better reward than a \emph{converged} MADDPG and independent DDPG. The dashed lines represent final rewards after convergence of the algorithms (i.e. no reward improvement after many episodes), and the solid red line represents decentralization of the expert. This demonstrates that we are able to successfully decentralize expert policies that achieve better rewards than a \emph{converged} MADDPG and independent DDPG. In other words, CESMA is able to find better optimum that MADDPG and independent DDPG were not able to find.}
	\label{fig:better_opt}
\end{figure}

\subsection{Cooperative Navigation with Communication}\label{subsec:coopcomm}

\edit{In these experiments, our experimental parameters are the same as in the Cooperative Navigation case above.} Here we adapt CESMA to a task that involves communication. In this scenario, the communication action taken by each agent at time step $t-1$ will appear as an observation to other agents at time step $t$. Although we require continuous communication to backprop, in practice we can use the softmax operator to provide the bridge between the discrete and continuous, as done in MADDPG (see the Gumbel-Softmax \cite{jang2016categorical}). And during decentralized execution, our agents are able to act with discrete communication inputs.

We examine two scenarios for CESMA that involve communication, and use the training scenario described in section \ref{subsec:cesma_comm}. The first scenario called the ``speaker and listener" environment has a speaker who broadcasts the correct goal landmark (in a ``language" it must learn) out of a possible 3 choices, and the listener, who is blind to the correct goal landmark, must use this information to move there. Communication is a necessity in this environment. The second scenario is cooperative navigation with communication and here we have three agents whose observation space includes the goal landmark of the \emph{other} agent(s), and not their own, and there are five possible goal landmarks. 

We see in Figure \ref{fig:learning_curves} that we achieve a higher reward in a more sample efficient manner. For the speaker and listener environment, \edit{using CESMA the decentralized multi-agents are able to immediately learn the correct communication protocol in order to solve the environment}. And MADDPG has a much higher variance in its convergence. 

We also see in Figure \ref{fig:better_opt} that the centralized expert was again able to find a policy that achieved a higher reward than a \emph{converged} MADDPG, and we were able to successfully decentralize this to obtain a decentralized multi-agent policy achieving the same superior reward as the expert.

\subsection{The issue with learning communication under decentralized learning}\label{subsec:decent_learning_comm}

\edit{It is worth noting in Figure \ref{fig:better_opt}, that the decentralized learner -- ind. DDPG -- has a hard time in the communicative environments, as compared to the centralized learning but decentralized execution methods of CESMA and MADDPG. In fact, it converges to a degenerate/wrong solution! This is because the learning of communication suffers from sparse rewards. The primary issue is that agents do not receive a reward signal on communication, and thus must figure out a protocol for themselves. This problem is exacerbated by the ambiguity of whether a bad reward outcome was due to the communications of a broadcasting agent, or the actions of the receiving agent. Also in Figure \ref{fig:better_opt}, we see MADDPG suffers a bit from this too as it converges to a lower reward than CESMA in the communicative environments. \\
\indent The advantage of CESMA is that there exists an expert action, and thus we have removed the ambiguity of whose fault it was that the multi-agent team received a bad reward -- we know which correct action to take. Furthermore, due to the structure of the neural networks (see Figure \ref{fig:cesma_with_comm}), we are able to backpropagate this expert error signal all the way to the weights of the broadcasting agent's communication nodes -- this means the broadcasting agents now has a direct signal for updating their weights in order to improve communication. In this way, we have accelerated learning the communication protocol by taking advantage of backpropagation.}

\subsection{DQNs and One-at-a-time Expert}\label{subsec:dqns-otex}

\begin{figure}[h]
	\begin{minipage}{\figscale\linewidth}
		\centering
		\includegraphics[width=\linewidth]{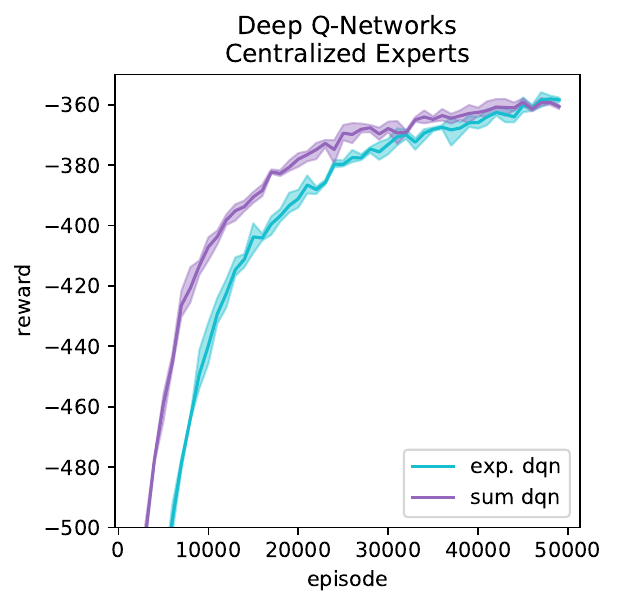}
	\end{minipage}
	\begin{minipage}{\figscale\linewidth}
		\centering
		\includegraphics[width=\linewidth]{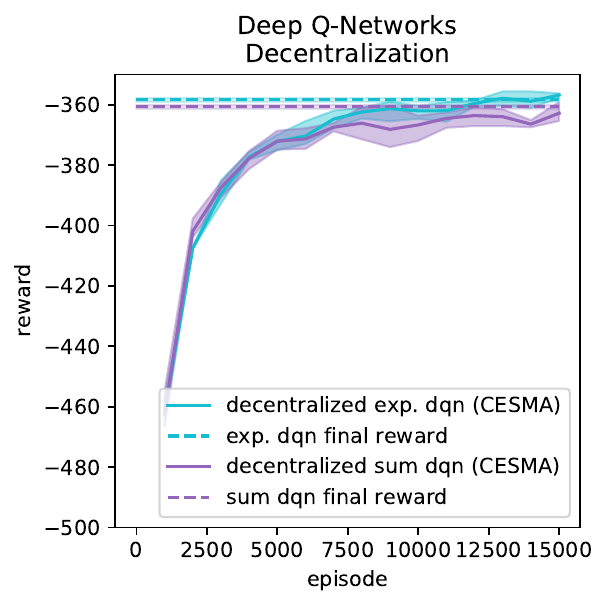}
	\end{minipage}%
	\begin{minipage}{\figscale\linewidth}
		\centering
		\includegraphics[width=0.94\linewidth]{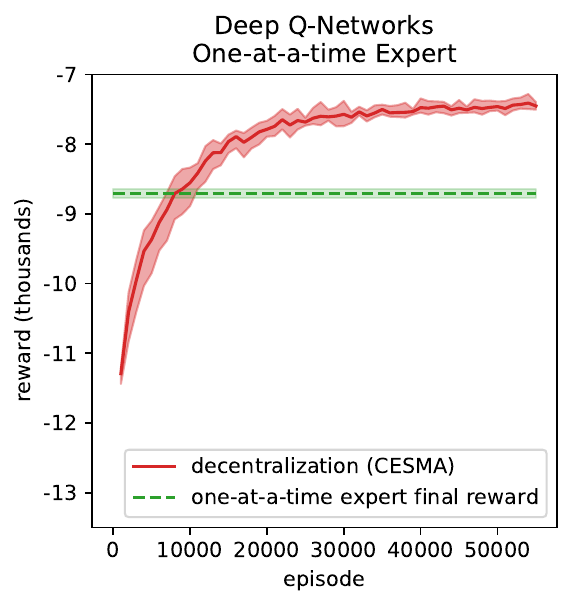}
	\end{minipage}
	\caption{Reward curves for decentralizing DQNs. The left graph shows reward curves for the exponential DQN and a centralized VDN (i.e. summing the Q-values). The middle graph shows the result of decentralizing these experts. The right graph shows the reward curves for decentralizing a centralized expert that can only move one agent at a time. We note the decentralized multi-agents achieve a better reward than the one-at-a-time centralized expert, because the agents have learned to move simultaneously.}
	\label{fig:DQNs}
\end{figure}

%%%%%%
% DQNs
%%%%%%

Here we examine decentralizing DQNs. We note further details of hyperparameters and descriptions can be found in the appendix. 

In the first set of experiments, we used the cross entropy loss for supervised learning, and used the cooperative navigation environment with 3 nonhomogenous agents. Here we examined the exponential actions DQN, which is just a naive implementation of DQNs for the multi-agents, and a Centralized VDN/sum-of-Q-values DQN where the system $Q$-values are the sum of the individual agent $Q$-values (see Section \ref{subsec:cursedim}). Reward curves can be found in Figure~\ref{fig:DQNs} (left-most and middle).  As can be seen, both the exponential and sum-of-Q-values DQNs are able to successfully be decentralized.

In a second experiment, we examine decentralizing a centralized expert that is restricted to only move one agent at a time, while others ``do nothing" (one may say it's a ``one-at-a-time" expert). This is another strategy to reduce the exploding number of actions as the number of agents increase -- we now have linear growth of actions as the number of agents grows. We note this dimensionality reduction has better convergence guarantees than the sum-of-Q-values approach (see Section \ref{subsec:cursedim} for a discussion). The experiments are done with six homogeneous agents (without communication). In Figure \ref{fig:DQNs} (right-most), we see that the  agents are able to achieve a better reward than the centralized expert. Examining the motion of the decentralized agents, we found they have learned to move simultaneously. This is an interesting technique in the case where we want decentralized multi-agents that move simultaneously, but we don't have enough computational resources to find a centralized expert that moves agents simultaneously (because of the exploding action space). So we do not have to spend as many computational resources for learning the centralized expert by training a one-at-a-time expert, and when decentralizing we can leverage the natural inclination of decentralized multi-agents to move simultaneously. 

\section{Conclusion}

We propose a MARL algorithm, called Centralized Expert Supervises Multiagents (CESMA), which takes the popular training paradigm of centralized training, but decentralized execution. The algorithm first trains a centralized expert policy, and then adapts DAgger to obtain decentralized policies that execute in a decentralized fashion. 
We also formulated an approach that enables multi-agents to learn a communication protocol, \edit{which is notoriously hard for decentralized agents to learn}.
Experiments in a variety of tasks show that CESMA can train successful decentralized multi-agent polices at a low sample complexity 
Notably, the decentralization protocol often is able to achieve the same levels of cumulative reward as a centralized controller, which in our experiments often achieves higher rewards than the competing methods MADDPG and independent DDPG. \edit{And in particular, we demonstrate successful decentralization in even complex tasks such as StarCraft 2, where decentralized learners had a tough time learning.}

\section*{Acknowledgements}

A.T. Lin and S. Osher were partially supported by AFOSR MURI FA9550-18-0502. G. Mont\'ufar has received funding from the European Research Council (ERC) under the European Union’s Horizon 2020 research and innovation programme (grant agreement no 757983).
\bibliography{references}

% %%%%%%%%%%%%%%%%%%%%%%%%%%%%%%%%%%%%%%%%%%%%%%%%%% %
%                                                    %
% Appendix, move to a new document before submitting %
%                                                    %
% %%%%%%%%%%%%%%%%%%%%%%%%%%%%%%%%%%%%%%%%%%%%%%%%%% %

\onecolumn
\newpage
\appendix

\begin{center}
	\huge Appendix
	
	\hrulefill
\end{center}

\section{More Experiments}

All experiments, with the exception of section \ref{subsec:rew_v_loss_slow_fast}, were conducted over three runs, following the example of \cite{maddpg}. And all experiments, with the exception of when using DQNs, had the same hyperparameters as in the main paper (see Appendix \ref{sec:hyperparameters} for the hyperparameters).

\subsection{Experiment where each agent has its own dataset of trajectories}
\label{subsec:individual_dataset}

\begin{figure}[H]
	\centering
	\includegraphics[scale=0.5]{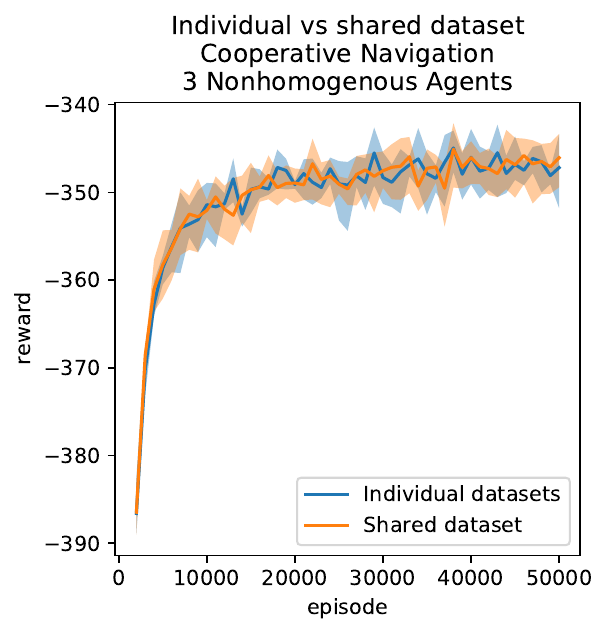}
	\caption{Reward curves where each agent has its own dataset of trajectories it learns from.}
\end{figure}

Here we describe an experiment where each agent has its own individual dataset of trajectories, versus a shared dataset. Namely, we plot the learning curves for decentralizing a policy in the two cases: (1) When each agent has its own dataset of trajectories, or (2) when there is a shared dataset of trajectories (which is the one we use in the experiments of the paper). We tested on the cooperative navigation environment with 3 nonhomogeneous agents. We hypothesized that the nonhomogeneity of the agents would have an effect on the shared reward, but this turned out not to be so. But it is interesting to note that in the main text, we found that the some agents had a bigger loss when doing supervised learning from the expert.

\subsection{Reward vs.\ loss, and slow and fast learners}
\label{subsec:rew_v_loss_slow_fast}

\begin{figure}[H]
	\begin{minipage}{0.24\textwidth}
		\centering
		\includegraphics[width=\textwidth]{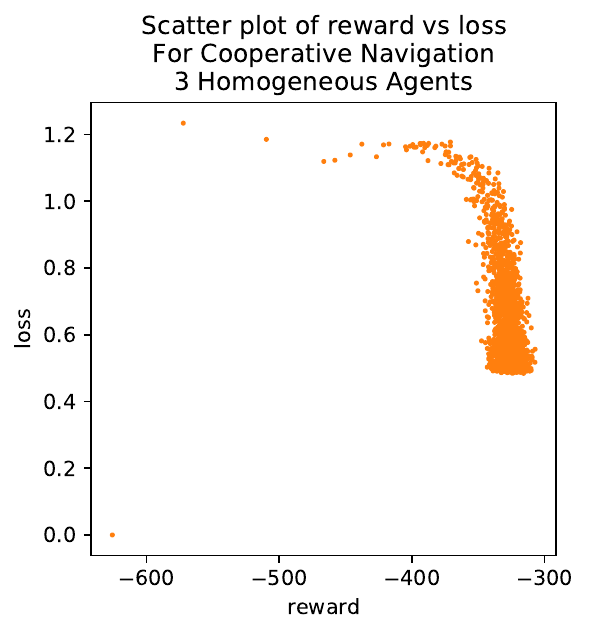}
	\end{minipage}
	\begin{minipage}{0.24\textwidth}
		\centering
		\includegraphics[width=\textwidth]{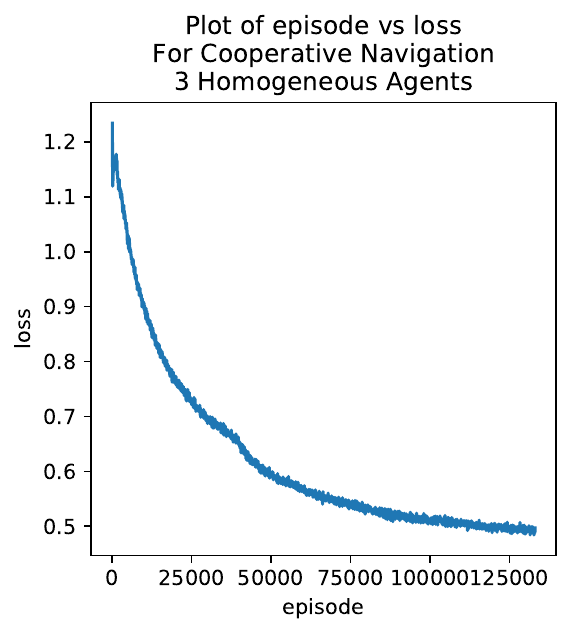}
	\end{minipage}
	\begin{minipage}{0.24\textwidth}
		\centering
		\includegraphics[width=\textwidth]{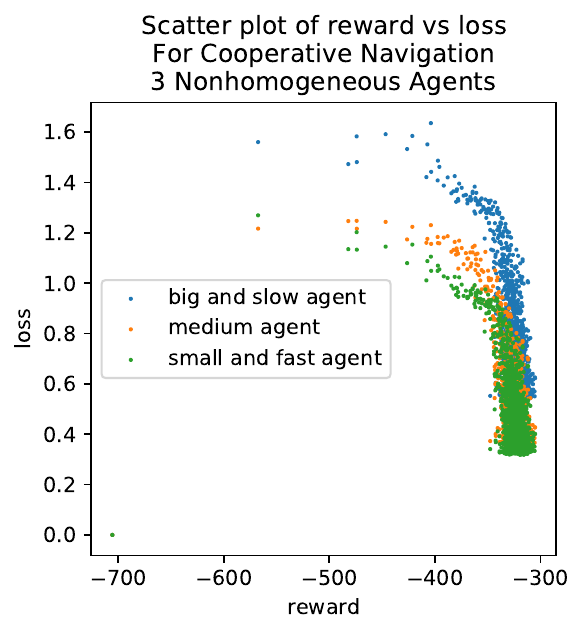}
	\end{minipage}
	\begin{minipage}{0.24\textwidth}
		\centering
		\includegraphics[width=\textwidth]{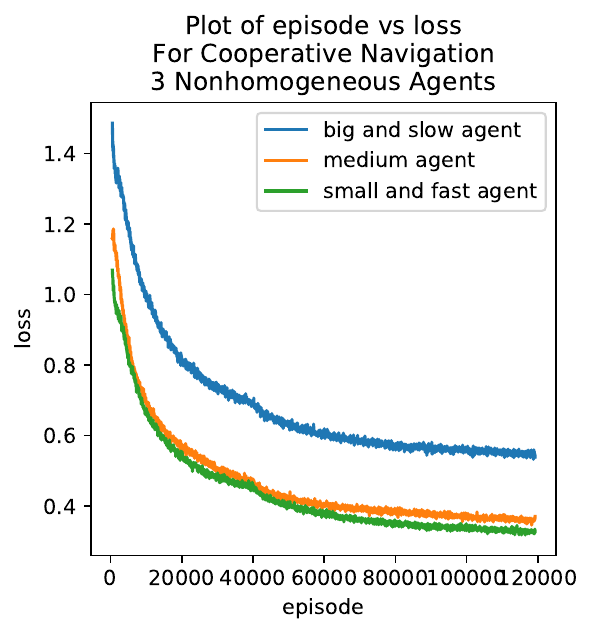}
	\end{minipage}
	\caption{Reward vs.\ loss, and loss vs.\ episode.}
	\label{fig:loss_v_reward}
\end{figure}

In our experiments with cooperative navigation, we reran the experiments in a truer DDPG fashion by solving a continuous version of the environment, and used the mean-squared error for the supervised learning. We examined the loss in the cooperative navigation task with 3 agents, both homogeneous and nonhomogeneous agents. We plot the figures in Figure \ref{fig:loss_v_reward}. We found that in these cases, the reward and loss were negatively correlated as expected, namely that we achieved a higher reward as the loss decreased. In the nonhomogeneous case, we plot each individual agents' reward vs its loss and found that the big and slow agent had the biggest loss, followed by the medium agent, and the small and fast agent being the quickest learner. This example demonstrates that in nonhomogeneous settings, some agents may be slower to imitate the expert than others.

We also observe that there is a decrease in marginal reward vs loss -- that is, at a certain point, one needs to obtain a much lower loss for a diminishing gains in reward. The hyperparameters are the same as in the main paper, described in section Appendix \ref{sec:hyperparameters}.

\pagebreak
\section{Pseudo-algorithm of CESMA (without communication)}
\label{sec:pseudocodeCESMAwithout}

Algorithm \ref{alg:supervising_multiagents} gives detailed pseudocode for CESMA without communicating agents. The notation follows the main paper.

\begin{algorithm}[H]
	\caption{CESMA: Centralized Expert Supervises Multi-Agents}
	\label{alg:supervising_multiagents}
	\begin{algorithmic}[1]
		\REQUIRE{A centralized policy $\pi^*$ that sufficiently solves the environment.}
		\REQUIRE{$M$ agents $\pi_{1}, \ldots \pi_{M}$, observation buffer $\mathcal{D}$ for multi-agent observations, batch size $B$}
		\WHILE{$\pi_{1}, \ldots, \pi_{M}$ not converged}
		\STATE{Obtain observations $o_1, \ldots, o_M$ from the environment}
		\STATE{Obtain agents' actions, $a_1 = \pi_{1}(o_1), \ldots, a_M = \pi_{M}(o_M)$}
		\STATE{Obtain expert action labels ${a}_i^* = \pi^*(o_1,\ldots,o_M)_i$, for $i=1,\ldots,M$}
		\STATE{Store the joint observation with expert action labels $((o_1, {a}_1^*),\ldots,(o_M, {a}_M^*)$ in $\mathcal{D}$}
		\IF{$|\mathcal{D}|$ sufficiently large}	
		\STATE{Sample a batch of $B$ multi-agent observations $\{((o^{(\beta)}_1, {a}_i^{*(\beta)}), \ldots, (o^b_M, {a}_M^{*(\beta)}))\}_{\beta=1}^B$}
		\STATE{Obtain $\pi^\text{new}_i$ by performing supervised learning for $\pi_{i}$ where the observation-label pairs are  $\{(o^{(\beta)}_i, a_i^{*(\beta)})\}_{\beta=1}^B$.}
		\STATE{$\pi_i \leftarrow \pi^\text{new}_i$}
		%			\STATE{$\pi_i \leftarrow \pi_i$ for $i=1,\ldots, M$}
		\ENDIF
		\ENDWHILE
	\end{algorithmic}
\end{algorithm}

\section{Pseudo-code of CESMA with communicating agents}

Algorithm \ref{alg:supervising_multiagents_comm} gives detailed pseudocode for CESMA with communicating agents. The notation follows the main paper.

\begin{algorithm}[H]
	\caption{CESMA: Centralized Expert Supervises Multi-Agents (Communicating Agents)}
	\label{alg:supervising_multiagents_comm}
	\begin{algorithmic}[1]
		\REQUIRE{A centralized policy $\pi^*$ that sufficiently solves the environment.}
		\REQUIRE{$M$ initial agents $\pi_{1}, \ldots \pi_{M}$, observation buffer $\mathcal{D}$ for multi-agent observations, batch size $B$}
		\REQUIRE{$\ell$, the supervised learning loss}
		\WHILE{$\pi_1, \ldots, \pi_M$ not converged}
		\STATE{Obtain the observations and communications $\{(o_i, c_i)\}_{i=1}^M$ from the environment.}
		\STATE{With these observations, obtain actions and step through the environment, to get new observations $\{\hat{o}_i\}_{i=1}^M$.}
		\STATE{Store the physical and communication observations together along with the expert label $\big(((o_1, c_1), \hat{o}_1, \hat{a}_1^*), \ldots, ((o_M, c_M), \hat{o}_M, \hat{a}_M^*)\big)$ in $\mathcal{D}$, where $\hat{a}^*_i = \pi^*(\hat{o}_1,\ldots, \hat{o}_M)_i$.}
		\IF{$|\mathcal{D}|$ sufficiently large}	
		\STATE{Sample a batch of $B$ multi-agent observations $\{((o_{1}^{(\beta)}, c_{1}^{(\beta)}),\hat{o}_{1}^{(\beta)}, \hat{a}_{1}^{*,(\beta)}), $ $\ldots, ((o_{M}^{(\beta)}, c_{M}^{(\beta)}), \hat{o}_{M}^{(\beta)}, \hat{a}_{M}^{*,(\beta)})\}_{\beta=1}^B$}
		\STATE{Obtain the up-to-date communication actions from each agent: $b^{(\beta)'}_{k} = \pi_k(o_{k}^{(\beta)}, c_{k}^{(\beta)})_{\text{comm}}$}
		\FOR{each agent $i=1$ to $M$}
		\STATE{\textbf{Communication loss:}}
		\STATE{For each agent $j\neq i$, obtain the up-to-date communication $\hat{c}_{j}^{(\beta)}$, which contains agent $i$'s communication action to agent $j$, so we can backprop to agent $i$'s weights}
		\STATE{Obtain the communication loss,
			\begin{equation*}
			\text{communication loss} = \frac{1}{B} \sum_{\beta=1}^B \frac{1}{M-1}\sum_{j=1, j\neq i}^M \ell(\hat{a}_j^{*,(\beta)}, \;\; \pi_{j}(\hat{o}_{j}^{(\beta)}, \;\hat{c}_{j}^{(\beta)'})_{\text{action}})
			\end{equation*}
			where the subscript ``action" denotes the physical action (and not the communication action), and where
			\begin{equation*}
			\hat{c}_j^{(\beta)'} = (b_{1}^{(\beta)'}, \ldots, \pi_i(o_i^{(\beta)}, c_i^{(\beta)}), \ldots, b_{j-1}^{(\beta)'}, b_{j+1}^{(\beta)'}, \ldots, b_{M}^{(\beta)'})
			\end{equation*}
		}
		\STATE{\textbf{Action loss:}}
		\STATE{Obtain the action loss:
			\begin{equation*}
			\text{action loss} = \frac{1}{B} \sum_{\beta=1}^B \ell(\hat{a}_i^{*,(\beta)}, \;\; \pi_{i}(\hat{o}_{i}^{(\beta)}, \;\hat{c}_{i}^{(\beta)} ))_{\text{action}})
			\end{equation*}
			where the subscript ``action" denotes the physical action (and not the communication action), and where,
			\begin{equation*}
			\hat{c}_i^{(\beta)'} = (b_{1}^{(\beta)'},\ldots, b_{i-1}^{(\beta)'}, b_{i+1}^{(\beta)'}, \ldots, b_{M}^{(\beta)'})
			\end{equation*}
		}
		\STATE{\textbf{Update:}}
		\STATE{Update the weights of $\pi_i$ where the total loss equals the action loss plus the communication loss, to obtain $\pi^{\text{new}}_i$.}
		\ENDFOR
		\STATE{Set $\pi_i \leftarrow \pi^{\text{new}}_i$, for $i=1, \ldots, M$.}
		\ENDIF
		\ENDWHILE
	\end{algorithmic}
\end{algorithm}

\section{Environments used in the experiments}

\subsection{StarCraft 2}\label{app:sc2}
\edit{
We use the StarCraft 2 environment \cite{samvelyan2019starcraft} in order to have a complex environment in which to test the effectiveness of CESMA against decentralized learning. In this environment, there are 8 allied units and 8 enemy units, and the allied units are the multi-agents whose policies should be learned, while the enemy units are controlled by AI created by the game developers, under the ``very hard" difficulty setting -- the hardest one before allowing the enemy to cheat. Each unit takes in the following properties of each allied and enemy unit: distance, relative x, relative y, health, shield, and unit type. Shields are an additional source of protection before damage to health can be dealt. The action space consists of a move action (4 directions), an attack option (for each enemy unit), a stop, and a no-op (dead agents can only choose no-op). Agents also have a sight-range/limited visibility, so this creates a partially-observable environment for each agent. The ultimate goal is to maximize the win rate, but rewards are given based on hit-point damage dealt, enemy units killed, and a special bonus for winning the battle.
}
\edit{
During each episode, the agents are given local observations at each timestep. If an agent dies, then the only action allowable for this agent is no-op. We note that each episode has varying timesteps, and our learning curve Figure \ref{fig:sc2_learning_curves} plots median test battle-win percentage vs timesteps, as done in \cite{samvelyan2019starcraft}. And we in no way change the environment defaults.
}
\edit{
Here our scenario consists of an 8 vs 8, where each team has 3 Stalker units and 5 Zealots units. An image of the environment is shown in Figure \ref{fig:sc2}. 
}

\edit{
In order to be perfectly fair, we follow the comparison procedure of \cite{samvelyan2019starcraft} and use their implementation of Independent Q-Learning (IQL). Thus, we perform five independent runs of IQL and CESMA, and compute the median win percentages. For both methods, the decentralized multi-agents are RNNs with 64 hidden units that only receive local observations. For the centralized expert of CESMA, we use the same architecture except now the input is the concatenated observations of all agents, and the outputs are the Q-values for each agent, and we take the sum of these Q-values to construct the expert Q-value -- this is well justified in \cite{sunehag2017value, qmix}. After training the centralized expert for 2 million timesteps, we choose the expert with the best reward to decentralize.
}

\subsection{Cooperative navigation}

The goal of this scenario is to have $N$ agents occupy $N$ landmarks in a 2D plane, and the agents are either homogeneous or heterogeneous. The environment consists of:
\begin{itemize}
	\item Observations: The (continuous) observations of each agent are the relative positions of other agents, the relative positions of each landmark, and its own velocity. Agents do not have access to other's velocities, and thus each agent only \emph{partially observes} the environment (aside from not knowing other agents' policies).
	\item Reward: At each timestep, if $A_i$ is the $i$th agent, and $L_j$ the $j$th landmark, then the reward $r_t$ at time $t$ is,
	\begin{align*}
	r_t = -\sum_{j=1}^N \min  \left\{ \| A_i - L_j \| : i=1,\ldots, N \right\}
	\end{align*}
	This is a sum over each landmark of the minimum agent distance to the landmark. Agents also receive a reward of $-1$ at each timestep that there is a collision.
	\item Actions: Each agents' actions are discrete and consist of: up, down, left, right, and do nothing. These actions are acceleration vectors (except do nothing), which the environment will take and simulate the agents' movements using basic physics (i.e. Newton's law).
\end{itemize}

\begin{figure}[H]
	\centering
	\includegraphics[scale=0.25]{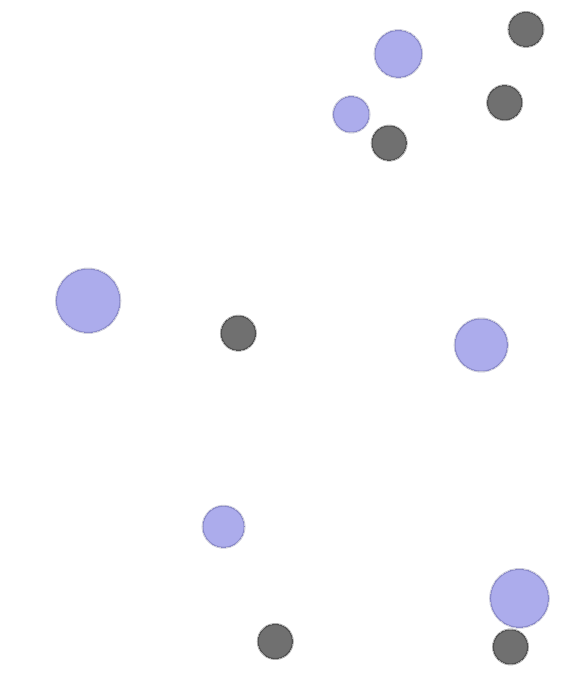}
	\caption{Example of cooperative navigation environment with 6 nonhomogeneous agents. The agents (blue) must decide how best to cover each landmark (grey).}
\end{figure}

\subsection{Speaker listener}

In this scenario, the goal is for the listener agent to reach a goal landmark, but it does not know which is the goal landmark. Thus it is reliant on the speaker agent to provide the correct goal landmark. The observation of the speaker is just the color of the goal landmark, while the observation of the listener is the relative positions of the landmark. The reward is the distance from the landmark.

\begin{itemize}
	\item Observations: The observation of the speaker is the goal landmark. The observation of the listener is the communication from the speaker, as well as the relative positions of each goal landmark.
	\item Reward: The reward is merely the negative (squared) distance from the listener to the goal landmark.
	\item Actions: The actions of the speaker is just a communication, a 3-dimensional vector. The actions of the listener are the five actions: up, down, left, right, and do nothing.
\end{itemize}

\begin{figure}[H]
	\centering
	\includegraphics[scale=0.25]{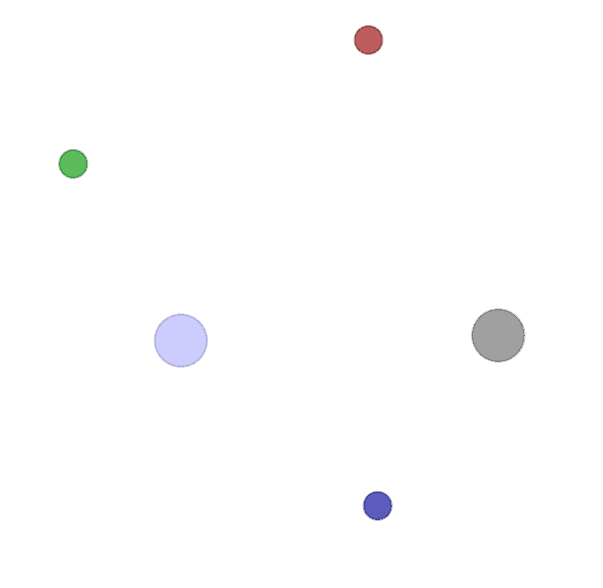}
	\caption{Example of the speaker and listener environment. The speaker (grey) must communicate to the agent which colored landmark to go towards (blue in this case).}
\end{figure}

\subsection{Cooperative navigation with communication}

In this particular scenario, we have one version with 2 agents and 3 landmarks, and another version with 3 agents and 5 landmarks. Each agent has a goal landmark that is only known by the other agents. Thus the each agent must communicate to the other agents its goal. The environment consists of:
\begin{itemize}
	\item Observations: The observations of each agent consist of the agent's personal velocity, the relative position of each landmark, the goal landmark for the other agent (an 3-dimensional RGB color value), and a communication observation from the other agent.
	\item Reward: At each timestep, the reward is the sum of the distances between and agent and its goal landmark.
	\item Actions: This time, agents have a movement action and a communication action. The movement action consists of either not doing anything, or outputting an acceleration vector of magnitude one in the direction of up, down, left, or right; so do nothing, up, down, left right. The communication action is a one-hot vector; here we choose the communication action to be a 10-dimensional one-hot vector. 
\end{itemize}

\begin{figure}[H]
	\centering
	\includegraphics[scale=0.25]{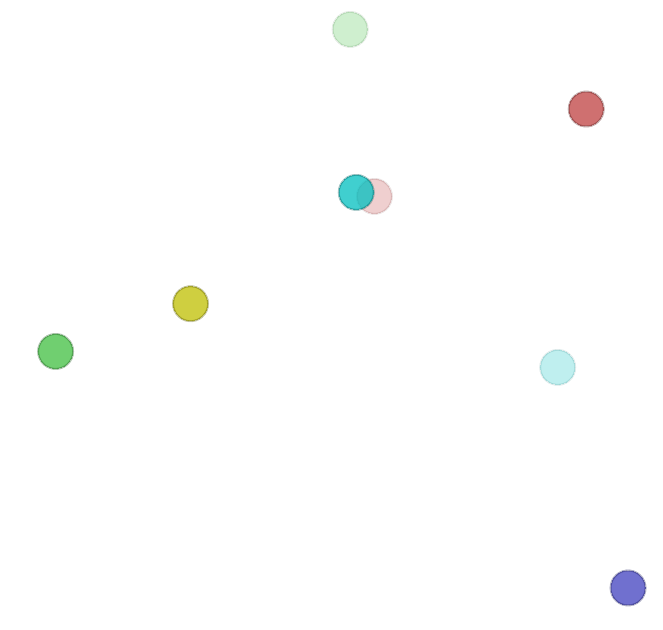}
	\caption{Example of cooperative navigation environment with communication. We have 3 agents and 5 landmarks. The lightly colored circles are agents and they must go towards their same-colored landmark.}
\end{figure}

\section{Hyperparameters}
\label{sec:hyperparameters}

\subsection{StarCraft 2}\label{app:sc2_hyp}

\edit{
For the Independent Q-Learner (IQL), we use the same hyperparameters in \cite{samvelyan2019starcraft}, namely this is an RNN with a fully-connected layer taking in the input, an RNN layer (GRUCell), and another fully-connected layer as output. The centralized expert and the decentralized multi-agents share the exact same architecture. For the multi-agents, again as in the aformentioned work, the parameters are shared among the multi-agents in order save computational cost, but each agent is give its own unit ID so the RNN is able to distinguish between heterogeneous units.
}

\edit{
For IQL, we train using Double Q-Learning trained after every episode, with a hard target update of every 200 episodes, optimized with RMSProp with a batch size of 32, a learning rate of 0.0005, a discount factor of 0.99, a gradient norm clipping of 10, and we use $\epsilon$-greedy action selection starting with $\epsilon=1$ which is annealed over 50$k$ episodes until it reaches its final $\epsilon$ of 0.05.
}

\edit{
For CESMA, the centralized expert is trained with Double Q-Learning using a soft target update with $\tau=0.001$, and optimized with the Adam optimizer, with a batch size of 32, a learning rate of 0.0002, weight decay of 0.0001, a gradient norm clipping of 10, and trained for every $20$ timesteps. Here the $\epsilon$-greedy action selection starts with 1 and is annealed for 3,000 episodes until its final $\epsilon$ of 0.05.
}

\edit{
For CESMA, the decentralized multi-agents are trained with supervised learning using the cross-entropy loss, with the labels being the one-hot actions of the centralized expert. We use the Adam optimizer with a batch size of 32, a learning rate of 0.0002, a weight decay of 0.0001, and we train every $20$ timesteps. Here, the agents only perform greedy action selection. 
}

\subsection{Comparing to MADDPG}

When we used DDPG to compare to MADDPG, our hyperparameters were:

\begin{itemize}
	
	\item For all environments, we chose the discount factor $\gamma$ to be $0.9$ for all experiments, as that seemed to benefit both the centralized expert as well as MADDPG (and as well as independently trained DDPG). And we always used a two-hidden-layer neural network for all of MADDPG's actors and critics, as well as the centralized expert, and the decentralized agents. The training of MADDPG used the hyperparameters from the MADDPG paper \cite{maddpg}, which we found to be quite optimal with the exception of having $\gamma=0.9$ (instead of 0.95), as that improved MADDPG's performance. In the graphs, the reward is averaged every 1,000 episodes.
	
	\item For the cooperative navigation environments with 3 agents, for both homogeneous and nonhomogeneous: Our centralized expert neural network was a two-hidden-layer neural network with 225 units each (as that matched the number of parameters for MADDPG when choosing 128 as their number of hidden units for each of their 3 agents), and we used a batch size of 64. The learning rate was 0.001, and $\tau=0.001$. We also clipped the gradient norms to $0.1$. When decentralizing, each agent was a two-hidden-layer neural network with 128 units (as in MADDPG), where we trained with a batch size of 32 and a learning rate of 0.001. In our experiment comparing with MADDPG, we use the cross entropy loss. The MADDPG and DDPG parameters were 128 hidden units, and we clipped gradients norms at 0.5, with a learning rate of 0.01.
	
	\item For the cooperative navigation with 6 agents, for both homogeneous and nonhomogeneous: Our centralized expert neural network was a two-hidden-layer neural network with 240 units each (as that matched the number of parameters for MADDPG when choosing 128 as their number of hidden units for each of their 3 agents' actor and critic), and we used a batch size of 32. The learning rate was 0.0001, and $\tau=0.0001$. We also clipped the gradient norms to $0.1$. When decentralizing, each agent was a two-hidden-layer neural network with 128 units (as in MADDPG), where we trained with a batch size of 32 and a learning rate of 0.001. In our experiment comparing with MADDPG, we use the cross entropy loss. The MADDPG and DDPG parameters were 128 hidden units, and we clipped gradients norms at 0.5, with a learning rate of 0.01.
	
	\item For the speaker and listener environment: Our centralized expert neural network was a two-hidden-layer neural network with 64 units each (which gave a lower number of parameters than MADDPG when choosing 64 as their number of hidden units for each of their 2 agents' actor and critic), and we used a batch size of 32. The learning rate was 0.0001, and $\tau=0.001$. When decentralizing, each agent was a two-hidden-layer neural network with 64 units (as in MADDPG), where we trained with a batch size of 32 and a learning rate of 0.001. In our experiment comparing with MADDPG, we use the cross entropy loss. The MADDPG and DDPG parameters were 64 hidden units, and we clipped gradients norms at 0.5, with a learning rate of 0.01.
	
	\item For the cooperative navigation with communication environment: Our centralized expert neural network was a two-hidden-layer neural network with 95 units each (which matched the number of parameters as MADDPG when choosing 64 as their number of hidden units for each of their 2 agents' actor and critic), and we used a batch size of 32. The learning rate was 0.0001, and $\tau=0.0001$. When decentralizing, each agent was a two-hidden-layer neural network with 64 units (as in MADDPG), where we trained with a batch size of 32 and a learning rate of 0.001. In our experiment comparing with MADDPG, we use the cross entropy loss. The MADDPG and DDPG parameters were 64 hidden units, and we clipped gradients norms at 0.5, with a learning rate of 0.01.
	
	\item We also run all the environments for 25 time steps.
	
\end{itemize}

\subsection{Deep Q-Networks (DQN)}

When we examined DQNs, our hyperparameters were:
	
\begin{itemize}
	\item DQNs: We used the cross entropy loss for the supervised learning portion, and used the cooperative navigation environment with 3 nonhomogenous agents. The DQNs we used are: the exponential actions DQN, which is just a naive implementation of DQNs for the multi-agents, and a Centralized VDN where the system $Q$ value is the sum of the individual agent $Q$ values. We used a neural network with 200 hidden units, batch size 64, and for the exponential DQN, we used a learning rate and $\tau$ of $5\times 10^{-4}$, and for the QMIX/Centralized VDN DQN we used a learning rate and $\tau$ of $10^{-3}$. We also used a noisy action selection for exploration. We stopped training of the decentralization once the mulit-agents reached the same reward as the expert; the dashed lines are a visual-aid that extrapolates the reward.
\end{itemize}

\subsection{One-At-A-Time Expert}

When we examined the one-at-a-time expert, the hyperparameters were:

\begin{itemize}
	\item The hyperparameters: (i) For the one-at-a-time expert, we used a fully-connected two-hidden-layer network with 380 units in each hidden layer, a batch size of 32, and a learning rate and $\tau$ of $10^{-4}$, and we used a gamma of 0.99, (ii) For decentralization, we used for each agent a fully-connected two-hidden-layer network with 128 units in each hidden layer, a batch size of 32, and a learning rate of $10^{-3}$. We also used a weighted cross-entropy loss function, where we gave more weight to an action; this makes sense as the observation database will naturally have much more do nothing actions, than a do something action, because the centralized expert only moves one agent at each timestep (and so the other multi-agent actions will be do nothing).
\end{itemize}

% ===============================================================================================
% How partial observability affects decentralization and the need for communication, LONG version
% ===============================================================================================

\section{The role of partial observability and communication}\label{sec:partobscomm}

Here we discuss more the role of partial observability and how it affects decentralization. We also comment on how communication can alleviate these issues.

\subsection{How partial observability affects decentralization}

In our setting of multi-agents, the centralized expert and the decentralized multi-agents have different structures of their policies, i.e.\ they are solving the problem in different policy spaces (this is desired, or else there would be little point in imitation learning, and we note the experiments in \cite{DAgger} also have this feature). The centralized expert observes the joint observations of all agents, and thus it is a function $\pi^*:\mathcal{O}_1 \times \cdots \times \mathcal{O}_M \rightarrow \mathcal{A}_1 \times \cdots \times \mathcal{A}_M$, and we can decompose $\pi^*$ into 
\begin{equation*}
\pi^*(\textbf{o}) = (\pi^*_1(\textbf{o}) , \ldots, \pi^*_M(\textbf{o})), 
\end{equation*}
where $\pi^*_i:\mathcal{O}_1 \times \cdots \times \mathcal{O}_M \rightarrow \mathcal{A}_i$. The goal of decentralization is to find multi-agent policies $\pi_1, \ldots, \pi_M$ such that 
\begin{equation*}
\pi^*(\textbf{o}) = (\pi^*_1(\textbf{o}) , \ldots, \pi^*_M(\textbf{o})) \overset{\text{want}}{=} (\pi_i(o_1), \ldots, \pi_M(o_M)) . 
\end{equation*}
Note that $\pi^*_i$ is able to observe the joint observations while $\pi_i$ is only able to observe its own local observation $o_i$. But from this constraint, this means we may encounter issues where
\begin{align*}
&\pi^*_i(o_1, \ldots, o_{i-1}, o_i, o_{i+1}, \ldots, o_M) = a_i, \\
&\quad \text{but} \quad \pi^*_i(\tilde{o}_1, \ldots, \tilde{o}_{i-1}, o_i, \tilde{o}_{i+1}, \ldots, \tilde{o}_M) = \tilde{a}_i, 
\end{align*}
so we want $\pi_i(o_i) = a_i \text{ or } \tilde{a}_i, \text{ or even something else}$. Thus the multi-agent policy can act sub-optimally in certain situations, being unaware of the global state. This unfortunate situation not only afflicts our algorithm, but any multi-agent training algorithm (and in general, any algorithm attempting to solve a POMDP). This is due to the partial observability problem in the multi-agent setting. More concretely, we can say partial observability is a problem for decentralization if there exists observations $(o_1, \ldots, o_{i-1}, o_i, o_{i+1}, \ldots, o_M)$, and $(\tilde{o}_1, \ldots, \tilde{o}_{i-1}, o_i, \tilde{o}_{i+1}, \ldots, \tilde{o}_M)$ such that 
\begin{align*}
&\pi^*(o_1, \ldots, o_{i-1}, o_i, o_{i+1}, \ldots, o_M) \\
&\qquad \neq \pi^*(\tilde{o}_1, \ldots, \tilde{o}_{i-1}, o_i, \tilde{o}_{i+1}, \ldots, \tilde{o}_M). 
\end{align*}

Relating this to the no-regret analysis in Theorem~\ref{thm:TCP}, the partial observability problem means that under certain environments it may be impossible for the multi-agents to match the expert exactly; this manifests in a cost $C_{p}$ where, 
\begin{align*}
\mu_N &= \min_{(\pi_1, \ldots, \pi_M)} \frac{1}{N} \sum_{i=1}^N \mathbb{E}_{\textbf{o}\,\sim\, d_{ (\pi^{(i)}_1, \ldots, \pi^{(i)}_M) }} [\ell(\textbf{o}, (\pi_1, \ldots, \pi_M)] \\ 
&\ge C_p , \quad \text{for all $N\ge 1$},
\end{align*}
which implies from Theorem \ref{thm:TCP} that the best guarantee of the reward for the multi-agents is $R(\hat{\pi}_1, \ldots, \hat{\pi}_M)$ $\ge R(\pi^*) - TC_p - O(1)$. 

\emph{The main takeaway:} In the original DAgger setting (i.e.\ the single-agent MDP setting), under reasonable assumptions on the distribution of states \citep[see][Section~4.2]{DAgger}, as $N\rightarrow \infty$ the cumulative reward of the learner can approximate the cumulative reward of the expert arbitrarily closely. Here when analyzing the multi-agent setting, we find that because $\mu_N \ge C_{p}$, then the no-regret analysis guarantees that after $O(T\log^k(T))$ updates we will find a multi-agent policy that obtains a cumulative reward that is within $C_p$ of the expert. In relation to this, in Appendix \ref{subsec:rew_v_loss_slow_fast}, we perform experiments and analyse the supervised learning loss versus the reward obtained by the multi-agents.

\subsection{The need for communication} \label{ref:theorycomms}

Decentralization without communication is most effective when all multi-agents can observe the full joint observation. Then from the perspective of each agent the only non-stationarity is from other agents' policies (which is alleviated by decentralization).

But when each agent only has local observations, then to avoid the partial observability problem in decentralization, there is an incentive to communicate. Namely, we want for the multi-agent policy $(\pi_1, \ldots, \pi_M)$%,
\begin{align*}
&\pi^*(\textbf{o}) = (\pi^*_1(\textbf{o}) , \ldots, \pi^*_M(\textbf{o})) \\
&\overset{\text{want}}{=} (\pi_i(o_1, c_1), \ldots, \pi_M(o_M, c_M)) , 
\end{align*}
where $c_i$ is the communication from either all or only some of the other agents, to agent $i$. Namely we view $c_i$ as a function $c_i:\mathcal{O}_1 \times \cdots \times \mathcal{O}_{i-1} \times \mathcal{O}_{i+1} \times \cdots \times \mathcal{O}_{M} \rightarrow \mathcal{C}_i$ (where $\mathcal{C}_i$ is some communication action space). Then we have the following requirement for the communication protocol $\{c_i\}_{i=1}$ in order to fix the partial observability problem in decentralization,
\begin{theorem}\label{thm:comm}
	If the multi-agent communication $c_i:\mathcal{O}_1 \times \cdots \times \mathcal{O}_{i-1} \times \mathcal{O}_{i+1} \times \cdots \times \mathcal{O}_{M} \rightarrow \mathcal{C}_i$ satisfies the %following 
	condition%:
	\begin{equation*}
	\begin{split}
	&\pi^*(o_1,\ldots, o_{i-1}, o_i, o_{i+1}, \ldots, o_M) \\ 
	&\quad \neq \pi^*(\tilde{o}_1,\ldots, \tilde{o}_{i-1}, o_i, \tilde{o}_{i+1}, \ldots, \tilde{o}_M), \\
	&\text{implies that }\\ 
	&c_i(o_1,\ldots, o_{i-1}, o_{i+1}, \ldots, o_M) \\
	&\quad \neq c_i(\tilde{o}_1,\ldots, \tilde{o}_{i-1}, \tilde{o}_{i+1}, \ldots, \tilde{o}_M)
	\end{split}
	\end{equation*}
	for all $i=1,\ldots, M$, then there is no cost due to partial observability in decentralization when the multi-agents use $\{c_i\}_{i=1}^M$ as their communication protocol, i.e.\ the multi-agents can match the expert perfectly on all observations.
\end{theorem}

\textbf{Remarks before proof:} The theorem above says that a sufficient condition for the communication protocol $\{c_i\}_{i=1}$ is that from the perspective of, say, agent $j$, then $c_j$ is able to provide information to agent $j$ about when the expert decides to output different actions for different global observations, even if the global observations share $o_j$ as a local observation.

Paired with Theorem 1, this implies that under the correct communication protocal, the multi-agents can approximate the expert arbitrarily closely (and that we need $O(T\log^k(T))$ updates). Of course, in our experiments we learn this communication protocol.
\begin{proof}
	
	This theorem is just intuitively saying that if the expert decides to choose different actions, i.e.
	\begin{align*}
	\pi^*(o_1,\ldots, o_{i-1}, o_i, o_{i+1}, \ldots, o_M) \neq \pi^*(\tilde{o}_1,\ldots, \tilde{o}_{i-1}, o_i, \tilde{o}_{i+1}, \ldots, \tilde{o}_M)
	\end{align*}
	then even though agent $i$ only sees $o_i$ in both cases, the communication $c_i$ to agent $i$ must be able to differentiate between them. Then we have the mathematical proof below:
	
	By assumption, for an agent $j$ with observations $\textbf{o} = (o_1,\ldots, o_{j-1}, o_j, o_{j+1}, \ldots, o_M)$ and $\tilde{\textbf{o}} = (\tilde{o}_1,\ldots, \tilde{o}_{j-1}, o_j \tilde{o}_{j+1}, \ldots, \tilde{o}_M)$ such that
	\begin{equation*}
	\pi^*(\textbf{o})_j = a_j, \quad \pi^*(\tilde{\textbf{o}})_j = \tilde{a}_j, \quad \text{but} \quad a_j \neq \tilde{a}_j.
	\end{equation*}
	then denoting $\textbf{o}_{-j}$ as the observation without $o_j$ and similarly for $\tilde{\textbf{o}}_{-j}$, then our assumption implies $c_j(\textbf{o}_{-j}) \neq c_j(\tilde{\textbf{o}}_{-j})$. Then clearly we can construct a policy where $\pi_j(o_j, c_j(\textbf{o}_j)) \neq \pi_j(o_j, c_j(\tilde{\textbf{o}}_j))$, because the inputs to $\pi_j$ are different.
	
	And so the multi-agents, using the communication protocol of $\{c_i\}_{i=1}^M$, can detect when an expert decides to change its action based on differences in the \emph{global}  observation (i.e. $\textbf{o}$ and $\tilde{\textbf{o}}$) even when the \emph{local} observation (i.e. $o_j$) stays the same.

\end{proof}

\end{document}